# scientific reports

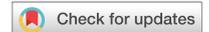

OPEN

# InvSim algorithm for pre-computing airplane flight controls in limited-range autonomous missions, and demonstration via double-roll maneuver of Mirage III fighters

Osama A. Marzouk

In this work, we start with a generic mathematical framework for the equations of motion (EOM) in flight mechanics with six degrees of freedom (6-DOF) for a general (not necessarily symmetric) fixed-wing aircraft. This mathematical framework incorporates (1) body axes (fixed in the airplane at its center of gravity), (2) inertial axes (fixed in the earth/ground at the take-off point), wind axes (aligned with the flight path/course), (3) spherical flight path angles (azimuth angle measured clockwise from the geographic north, and elevation angle measured above the horizon plane), and (4) spherical flight angles (angle of attack and sideslip angle). We then manipulate these equations of motion to derive a customized version suitable for inverse simulation flight mechanics, where a target flight trajectory is specified while a set of corresponding necessary flight controls to achieve that maneuver are predicted. We then present a numerical procedure for integrating the developed inverse simulation (InvSim) system in time; utilizing (1) symbolic mathematics, (2) explicit fourth-order Runge–Kutta (RK4) numerical integration technique, and (3) expressions based on the finite difference method (FDM); such that the four necessary control variables (engine thrust force, ailerons' deflection angle, elevators' deflection angle, and rudder's deflection angle) are computed as discrete values over the entire maneuver time, and these calculated control values enable the airplane to achieve the desired flight trajectory, which is specified by three inertial Cartesian coordinates of the airplane, in addition to the Euler's roll angle. We finally demonstrate the proposed numerical procedure of flight mechanics inverse simulation (InvSim) through an example case that is representative of the Mirage III family of French fighter airplanes, in which a straight subsonic flight with a double-roll maneuver over a duration of 30 s at an altitude of 5 km (3.107 mi or 16,404 ft) is inversely simulated.

Mathematical modeling and numerical simulations are important tools for describing various nonlinear complex phenomena and processes, as well as implementing computer-aided design (CAD), computational fluid dynamics (CFD), and automatic control[1–10]. Flight mechanics (also called flight dynamics) is one of the engineering fields that benefit largely from mathematical modeling and numerical simulation; because simple analytical reduced-order solutions in aerospace applications and aerodynamics are typically not available except under very restrained conditions; and experimental techniques through wind-tunnel tests (WTT) and flight tests are stochastic (non-deterministic), expensive, limited in terms of the amount of data that can be measured directly, and sometimes intrusive (influencing the domain being tested)[11–20].

Inverse simulation (InvSim) in flight mechanics is a normative category of flight mechanics modeling in which a desired flight trajectory (flight maneuver or flight mission) is specified through a number of inputs, while the corresponding flight controls (the model-based feedforward control variables) needed to achieve this trajectory

College of Engineering, University of Buraimi, Al Buraimi 512, Sultanate of Oman. email: osama.m@uob.edu.om

nature portfolio

are predicted; the opposite of this modeling process is called forward flight mechanics simulation, which is an exploratory category of flight mechanics[21–27]. Accurate inverse simulation flight mechanics modeling facilitates advanced modes of transportation through pilotless (autonomous) aviation activities, such as regular electrified urban air mobility (e-UAM) trips within smart cities or between neighbor cities, powered by clean renewables; although additional real-time control systems for external perturbation suppression should be augmented[28–32].

This work is a sequel to a previous part in which we provided a detailed mathematical framework for general modeling of the motion of a fixed-wing aircraft, and this framework has several advantages; namely: (1) all six degrees of freedom (6-DOF) are included, (2) the singularity of upward/downward vertical flight (encountered in a traditional Euler-angle representation) is avoided, (3) linear (translational) momentum equations are transformed such that the order of magnitude of adjacent terms does not suffer from large discrepancy under fast dynamics, (4) the assumption of airplane symmetry is eliminated, (5) the aerodynamic details for all the flight-dependent aerodynamic/stability coefficients are clearly expressed, (6) three sets of axes (namely, inertial ground/earth axes, body-fixed axes, and wind axes) are utilized to describe the attitude of the airplane, rather than using the Euler angles solely, (7) two flight-path angles (azimuth and elevation) are utilized as an intermediate spherical coordinate system; allowing the separation of the flight path direction relative to air (this is the airplane "course") from the airplane attitude relative to the ground (this is the airplane "heading"), (8) only scalar equations (rather than vector equations or quaternions) are utilized, which simplifies the implementation process as a computer-based flight mechanics simulator, (9) the variation of the air density with altitude is accounted for, using the international standard atmosphere (ISA) model for air as an ideal homogeneous gas, and (10) the model can be easily adjusted to specific airplane conditions through user-defined input parameters or specialized airplane features (such as a nonlinear lift coefficient profile, instead of linear dependence on the angle of attack) can be handled through minor modifications[33–56].

In the current work, the previously explained general nonlinear differential–algebraic equations (DAE) system for flight mechanics is further reformulated such that they fit specifically inverse simulation flight mechanics, and a detailed computational algorithm is presented to numerically integrate these equations of motion (EOM) through explicit mathematical expressions that do not require solving an algebraic system of equations, with the use of the fourth-order Runge–Kutta (RK4) numerical integration method; also numerical differentiation expressions based on the finite difference method (FDM) may be used[57–64].

We implement this proposed numerical procedure as a computer code, and demonstrate its utilization while inversely simulating a continuous-double-roll maneuver with a set of airplane data that nearly corresponds to the Mirage III fighter aircraft, produced by the French aerospace company "Dassault Aviation"[65,66].

## Research method
### Problem statement

The inverse simulation (InvSim) flight mechanics algorithm proposed here considers the airplane as a MIMO (multi-input multi-output) control system, with four inputs and four outputs. The four inputs are specified as the inertial (ground-referenced) Cartesian coordinates $(x_g, y_g, z_g)$ of the airplane's center of gravity (CG), and these coordinates may be described as analytical functions of time or as discrete values recorded at specified time values, uniformly spaced with a constant time step $(\Delta t)$. The $(x_g)$ coordinate represents the signed displacement traveled by the airplane in the geographic north direction from the initial flight point (the take-off point). The $(y_g)$ coordinate represents the signed displacement traveled by the airplane in the geographic east direction from the initial flight point. The $(z_g)$ coordinate represents the signed displacement traveled by the airplane toward the earth's center from the initial flight point, and thus this coordinate is expected to have negative values except during parts of the maneuver where the airplane descends to an altitude below the initial altitude. The three inertial (ground-referenced) Cartesian coordinates $(x_g, y_g, z_g)$ are expressed in meters (m). Figure 1 illustrates these three input coordinates. The term "gravity axis" here refers to the inertial axis pointing toward the earth's center (perpendicular to the horizon plane), and it is opposite to the "altitude axis" that is also perpendicular to the horizon plane but points toward the sky. The origin of these ground-referenced rectangular coordinates is the initial flight point (the first location of the trajectory to be inversely simulated).

In the current work, the altitude of the airplane's center of gravity as measured from the mean sea level (MSL) is designated by the symbol $(h)$, and the initial altitude is designated by the symbol $(h_{ini})$. Therefore, the gained height of the airplane's center of gravity above the initial trajectory point is $(h - h_{ini})$; which should be equal to the negative value of the $(z_g)$ coordinate. Therefore,

$$h - h_{ini} = -z_g \tag{1}$$

The fourth input to the proposed InvSim model is the roll angle (expressed in radians, "rad"), which is one of the three Euler angles, and it describes the lateral attitude of the airplane. If the starboard (right) tip and the port (left) tip of the wing are at the same vertical position (having the same altitude), then the roll angle $(\phi)$ is zero. By convention, the roll angle is positive if the wing's starboard/right tip tilts down and the wing's port/left tip tilts up[67,68]. Figure 2 illustrates the roll angle in a particular configuration where it has a positive value.

The four outputs from the proposed inverse simulation (InvSim) flight mechanics model here are the four flight controls (four control variables), computed as four discrete series of values (four numerical vectors). These controls are in charge of adjusting the speed and orientation of the airplane, such that the input conditions are satisfied. These four output controls are:

1. The engine thrust force $(T)$, in newtons (N). This control can also be an electric propulsion force in the case of using an electric propeller or an electric ducted fan (EDF) instead of a traditional fuel-fired heat engine[69–74], and this electric propulsion[75–81] (or the use of heat engines powered by clean alternative non-fossil

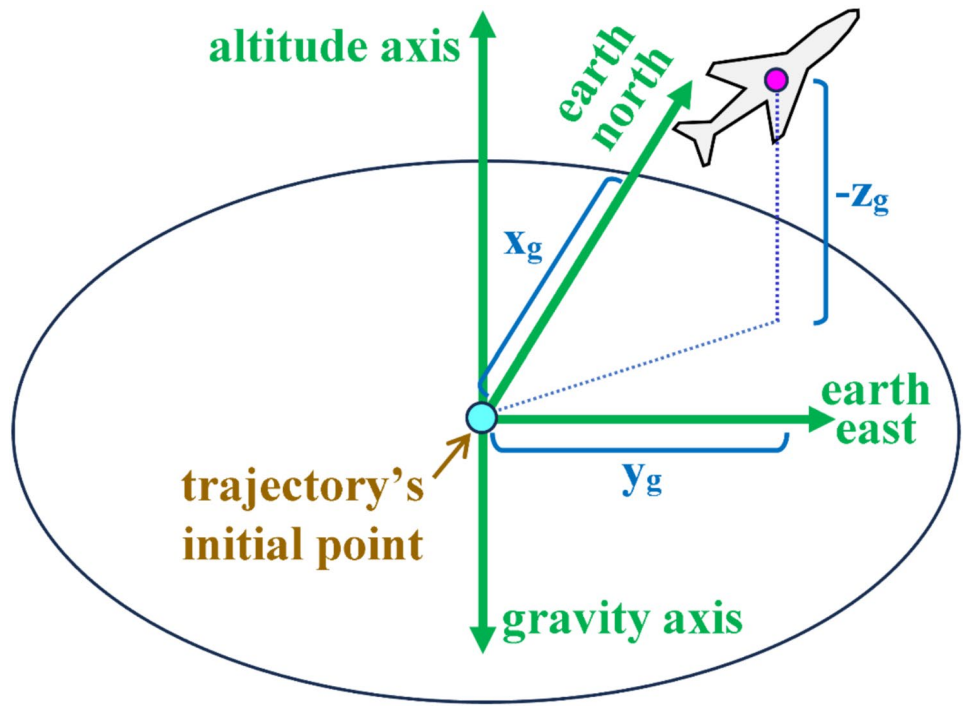

**Fig. 1.** Illustration of the three inertial (ground-referenced) coordinates, which are three of the four inputs to InvSim.

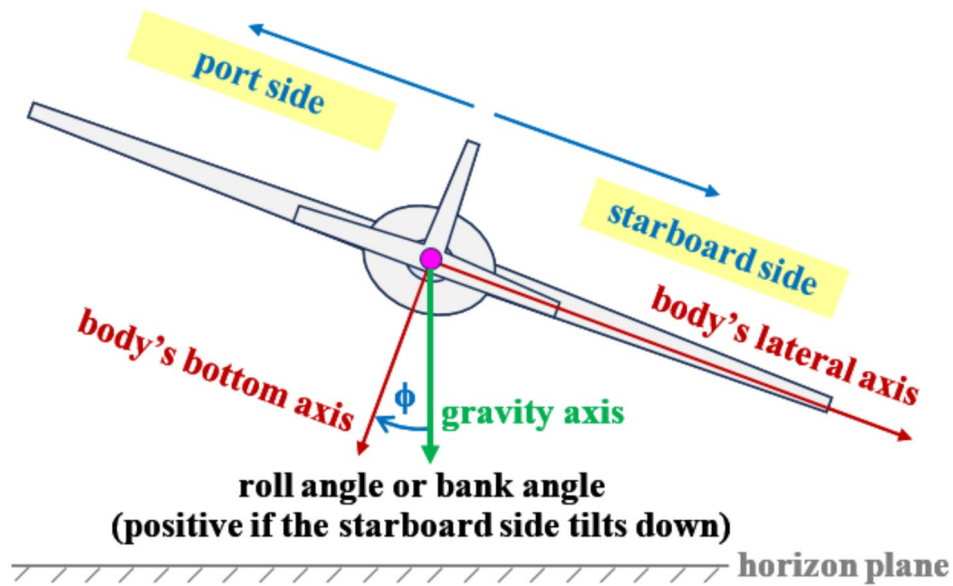

**Fig. 2.** Illustration of the roll angle, which is the fourth input to InvSim.

fuels[82–85]) has an environmental advantage due to eliminating greenhouse (GHG) emissions as a results of burning a fossil fuel[86–89]. While the presented InvSim model theoretically can admit thrust reversal (TR) or reverse thrust (corresponding to negative thrust values) during brief periods of the whole trajectory, this is typically not acceptable in fixed-wind airplanes[90–93].

2. The ailerons' deflection angle ($\delta_l$), in radians (rad). This deflection is primarily in charge of the roll degree of freedom. This deflection angle is positive when the hinged starboard/right aileron tilts up and simultaneously the hinged port/left aileron tilts down (which induces a rolling moment leading to a positive roll angle, $\phi$).

3. The elevators' deflection angle ($\delta_m$), in radians (rad). This deflection is primarily in charge of the pitch degree of freedom. This deflection angle is positive when both hinged elevators tilt down (which induces a pitching moment leading to a positive pitch angle, $\theta$, where the airplane's nose tilts up).

4. The rudder's deflection angle ($\delta_n$), in radians (rad). This deflection is primarily in charge of the yaw degree of freedom. This deflection angle is positive when the hinged rudder tilts toward the starboard/right side (which induces a yawing moment leading to a positive yaw angle "heading angle", $\psi$, where the airplane's nose tilts toward the starboard/right side).

The sign convention of the three control deflection angles ($\delta_l, \delta_m, \delta_n$) are positive when the resulting body-referenced rotations become related to their respective body axes (the longitudinal or forward axis, the lateral or starboard axis, and the bottom or third axis) through the corkscrew rule (also called the curl right-hand rule or the right-hand grip rule)[94,95].

Figure 3 illustrates these four flight controls (which are the four main outputs from the InvSim model).

## Research approach

In order to solve the stated problem in the previous subsection, symbolic mathematical manipulation is combined with computational methods to build and present here a proposed numerical algorithm for solving the inverse simulation flight mechanics problem. A general set of flight mechanics equations of motion are first presented. These equations are then reformulated to be in the inverse simulation InvSim form, such that the four output flight controls can be obtained for specified inputs (trajectory coordinates and roll angle). Then, a proposed algorithm is presented, which allows numerically integrating the reformulated coupled differential–algebraic equations (DAE) in a simple way that can be implemented using a general-purpose computer programming language, without the need for specialized capabilities or commercial packages[96–103]. We here use the MATLAB/Octave programming language, which is a high-level interpreted language suitable especially for numerical computations, although other programming languages (such as Python or FORTRAN) can also be used[104–111].

The success of the proposed computational InvSim algorithm and the underlying mathematical model is supported by numerically simulating an example flight maneuver. We obtained some symbolic expressions for needed derivative terms manually using normal calculus rules; and independently using Mathematica (a popular software tool for symbolic mathematics and computations that has been used in many research works before), and both sets of obtained expressions were exactly compatible[112–122].

## General equations of motion

In this section, we present part of a flight mechanics mathematical framework, which is not optimized specifically for inverse simulations. The equations of motion listed in this section are not just the six main equations (the three linear/translational momentum equations and the three angular/rotational momentum equations) for a six-degree-of-freedom rigid body in flight. Instead, auxiliary intermediate equations are also listed, since they need to be considered along with the six main equations as a complete integrated coupled system.

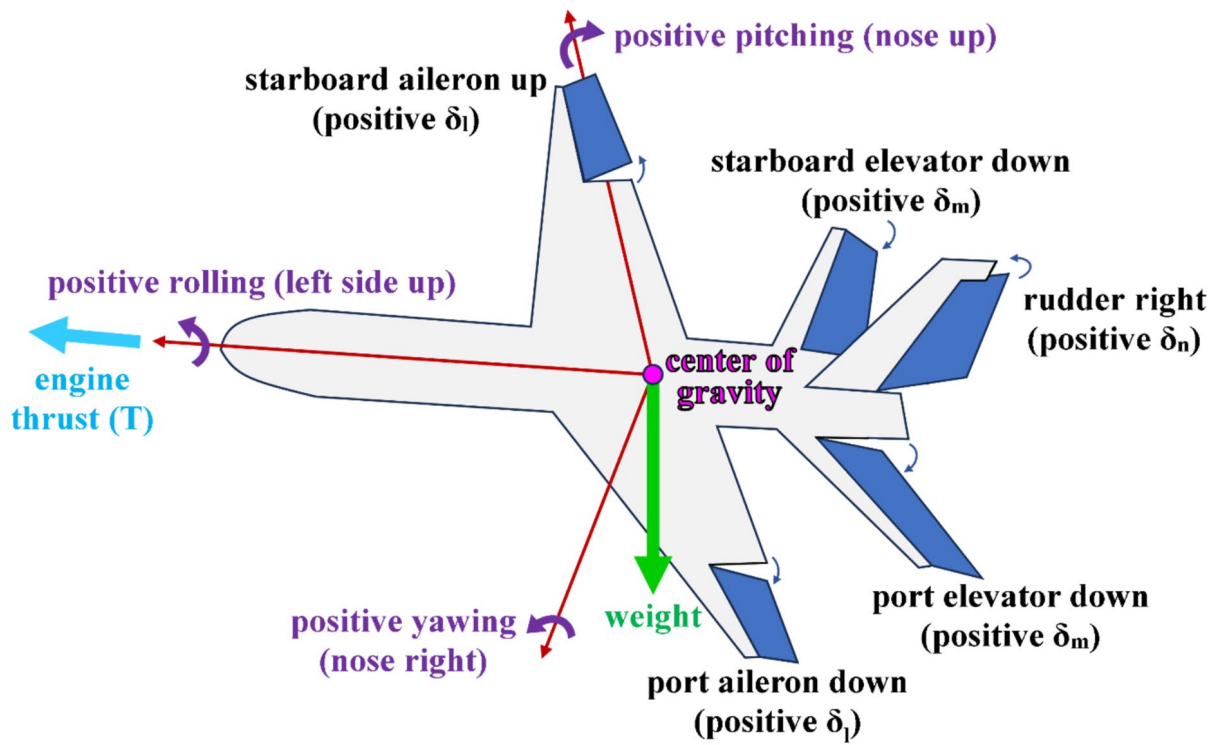

**Fig. 3.** Illustration of the four flight controls.

## Angular velocity vector in body axes

The roll rate ($p$), pitch rate ($q$), and yaw rate ($r$) are related to the Euler angular rates or Euler rates ($\dot{\phi}, \dot{\theta}, \dot{\psi}$) that are the time derivatives of the Euler angles, according to

$$p = \dot{\phi} - \sin\theta\dot{\psi} \tag{2}$$

$$q = \cos\phi\dot{\theta} + \cos\theta\sin\phi\dot{\psi} \tag{3}$$

$$r = \cos\theta\cos\phi\dot{\psi} - \sin\phi\dot{\theta} \tag{4}$$

Figure 4 illustrates the three Euler angles ($\phi, \theta, \psi$) which describe the three-stage rotational transformation from the inertial axes system (global/earth north, global/earth east, and global/earth gravity) into the body-fixed axes system (longitudinal, lateral, bottom; or $x_b, y_b, z_b$)[123-126].

## Linear-momentum equations and equilibrium

The linear-momentum equations are transformed from the Cartesian system into the translating spherical wind axes, whose origin is translating with the body (coincides with the airplane's center of gravity), and whose coordinates are the velocity magnitude ($V$), the sideslip angle ($\beta$), and the angle of attack (AoA or $\alpha$). These three variables ($V, \beta, \alpha$) can be viewed as spherical coordinates, with the velocity magnitude ($V$) being the radial coordinate. The forward wind axis $x_w$ is tangent to the flight path and thus coincides with the total velocity vector (this is the velocity of the airplane's center of gravity). The sideslip angle ($\beta$) represents a rotational transformation of the wind axes such that the rotated $x_w$ lies in the airplane's fixed plane $x_b - z_b$ (the airplane's midplane), which is the plane of symmetry for symmetric airplanes.

The condition ($\beta = 0$) means that the incoming airflow relative to the airplane is symmetric with respect to the airplane's midplane. The angle of attack ($\alpha$) represents a subsequent rotational transformation (after the $\beta$ rotational transformation) such that the twice-rotated wind axes ($x_w, y_w, z_w$) coincide with the body-fixed axes ($x_b, y_b, z_b$). The components of the velocity vector ($\vec{V}$) along the body-fixed axes ($x_b, y_b, z_b$) are ($u, v, w$), respectively. Figure 5 illustrates the wind angles ($\alpha$ and $\beta$), as well as the relation between the wind axes system ($x_w, y_w, z_w$) and the body axes system ($x_b, y_b, z_b$).

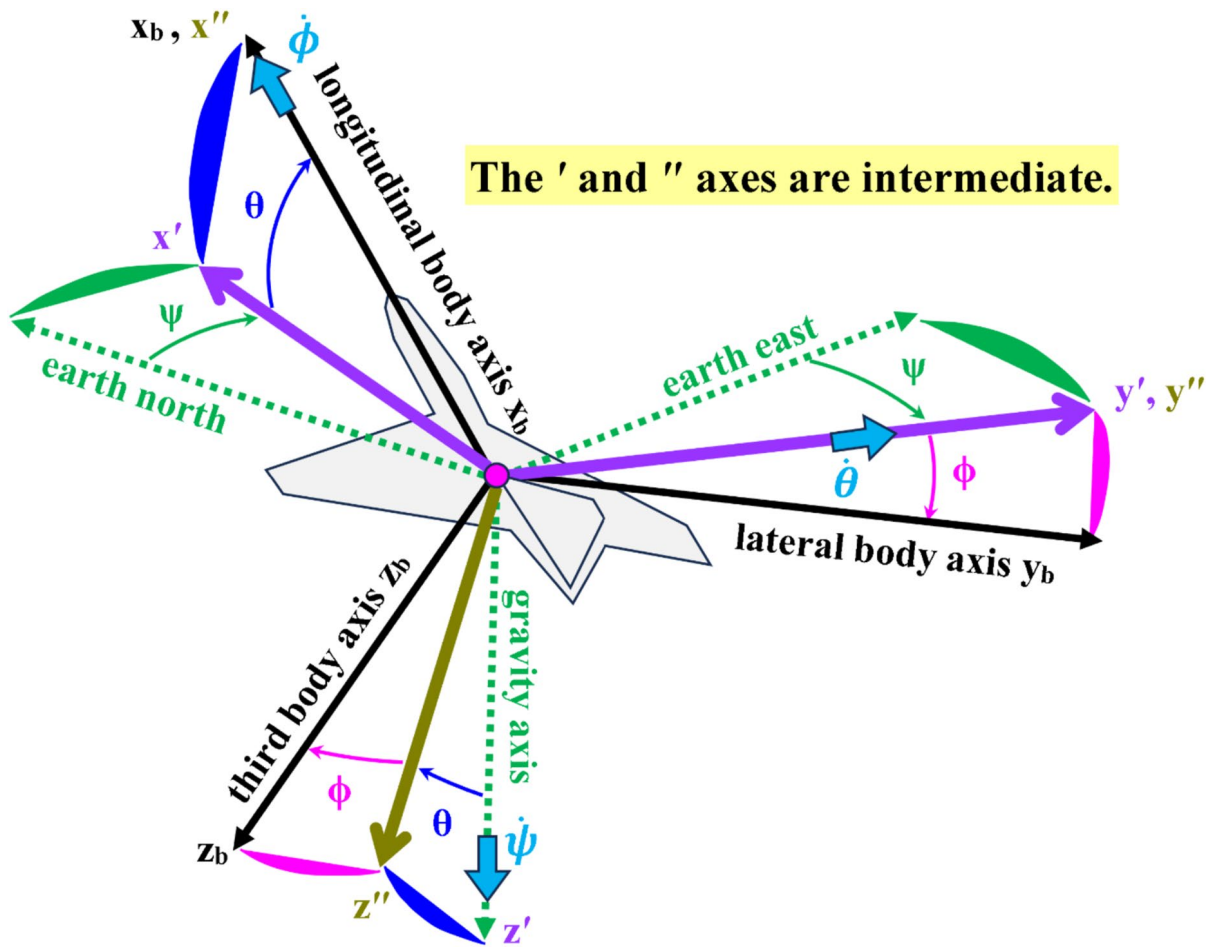

**Fig. 4.** Illustration of the three Euler angles, as well as the inertial axes and the body axes.

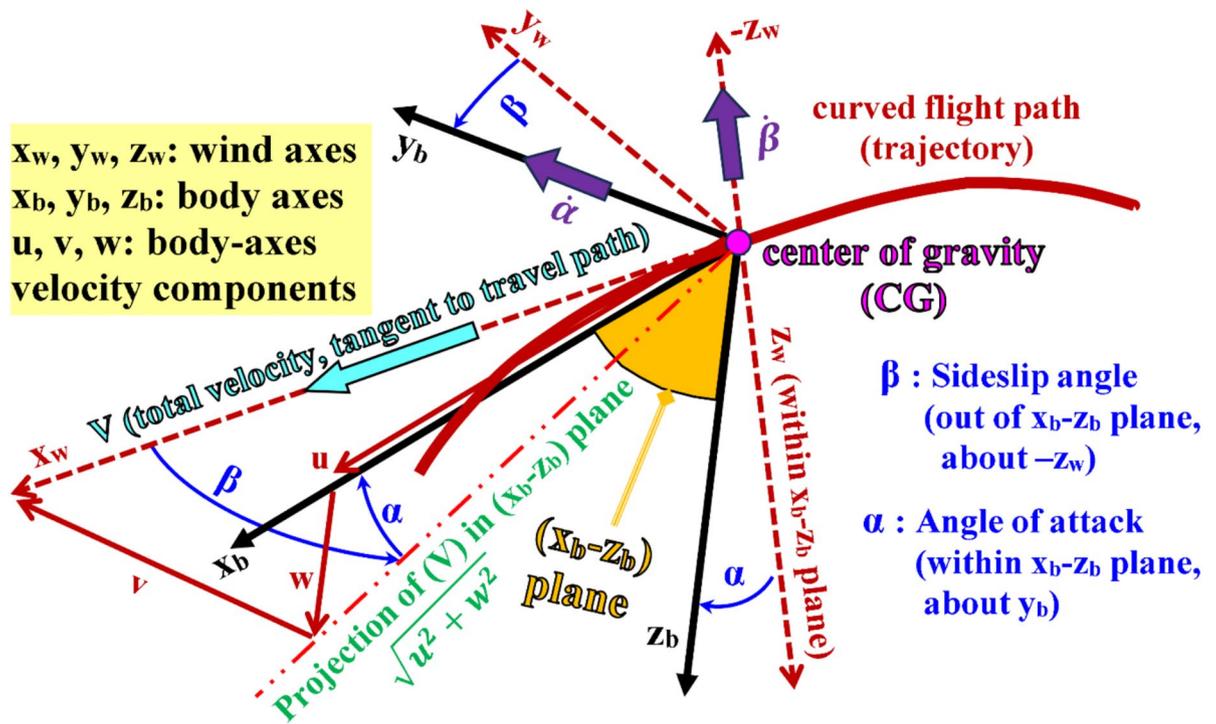

**Fig. 5.** Illustration of the wind axes and their angles (sideslip angle and angle of attack).

Although our definition for the angle of attack (AoA or $\alpha$) discussed above, as an angle of axes transformation, implies that in the absence of any sideslip ($\beta = 0$) and when $\alpha = 0$, the longitudinal body axis ($x_b$) coincides with the forward wind axis ($x_w$). This configuration corresponds to a horizontal steady-level flight with zero tilts by the airplane (the body-fixed plane $x_b - y_b$ is parallel to the horizon plane), and we refer to this configuration here as the "equilibrium" condition of flight. However, despite the horizontal orientation of the airplane's body, and the zero value assigned to the angle of attack ($\alpha$) as per our definition, aerodynamic principles demand that there must be an exerted lifting force to counteract the weight of the airplane during an equilibrium flight (as well as an exerted thrust force to counteract the drag force resisting the forward motion)[127–129].

The equilibrium lift force may require either (1) using a nonsymmetrical (cambered) wing airfoil section while keeping the wing mounted parallel to the main airplane body (the fuselage), such that a non-zero upward lifting force can be exerted even when the airplane is flying horizontally through the ambient air with its nose-to-tail longitudinal axis is also oriented horizontal ($x_w$); or (2) mounting the wing (which can be cambered or symmetric in this case, not necessarily cambered) at a small tilt angle called the angle of incidence or the mounting angle ($i$), such that when the airplane flies horizontally with a horizontal longitudinal body axis (a horizontal fuselage), there is still a non-zero lifting force that achieves the equilibrium condition because the tilted wing now faces the air asymmetrically, causing the air pressure at the lower surface of the wing to be higher than the air pressure at the upper surface of the wing due to the asymmetric air flow around the wing, and this leads to the upward lifting force component (combined with an aerodynamic drag force component)[130–143]. In the current study, we adopt the latter choice; thus, we assume that the wing is installed into the fuselage at a small angle of incidence ($i$) that satisfies the equilibrium condition; and this tilt angle ($i$) is excluded from the modeling and numerical simulation discussed here.

In the flight mechanics model presented here, the angle of attack ($\alpha$) excludes any equilibrium angle between the velocity vector of the airplane's center of gravity (whose magnitude is $V$) and the wing's mean chord line (the virtual straight line connecting the leading edge with the trailing edge of the wing of an airfoil section)[144–149]. Therefore, the symbol ($\alpha$) that appears in this study means the change in the conventional angle of attack (denoted here by the symbol $\widetilde{\alpha}$) from its equilibrium conventional value ($\widetilde{\alpha}_{equb}$), which is assumed to be equal to the angle of incidence in the current work. Therefore, we have

$$\alpha = \widetilde{\alpha} - \widetilde{\alpha}_{equb} \tag{5}$$

$$\widetilde{\alpha}_{equb} = i \tag{6}$$

$$\alpha = \widetilde{\alpha} - i \tag{7}$$

In aeronautics, the equilibrium conventional angle of attack ($\widetilde{\alpha}_{equb}$) is not truly a constant. Rather, this equilibrium conventional angle of attack depends on the flight speed, and it also depends on the air density, which in turn depends on the air pressure[150–152] (as for any ideal compressible gaseous medium such as ambient air[153–157]). This ambient air pressure drops with the flight altitude[158–160]. Therefore, our assumption of equality between the equilibrium conventional angle of attack and the angle of incidence ($\widetilde{\alpha}_{equb} = i$) implies that the angle of incidence is effectively adjustable; for example through using movable flaps attached to the wing, and these allows adjusting the wing's effective angle of incidence[161–163].

Figure 6 illustrates the difference between the angle of incidence ($i$) and the angle of attack ($\alpha$). The figure also shows how they differ from the Euler pitch angle ($\theta$) discussed in the previous subsection, and how they differ from the elevation flight path angle ($\theta_w$) that describes the climb angle of the airplane based on its course of flight as a point particle.

The linear-momentum equations (with their components corresponding to the wind axes) are

$$
\begin{aligned}
m\dot{V} = \ &\bar{q}S \left( C_x \ \cos\alpha\cos\beta + C_y \ \sin\beta + C_z \ \sin\alpha\cos\beta \right) \\
&+ mg \left( \cos\theta\sin\phi\sin\beta - \sin\theta\cos\alpha\cos\beta + \cos\theta\cos\phi\sin\alpha\cos\beta \right) \\
&+ T\cos\alpha\cos\beta
\end{aligned}
\tag{8}
$$

$$
\begin{aligned}
mV\dot{\beta} = \ &\bar{q}S \left( C_y \ \cos\beta - C_x \ \cos\alpha\sin\beta - C_z \ \sin\alpha\sin\beta \right) \\
&+ mg \left( \cos\theta\sin\phi\cos\beta + \sin\theta\cos\alpha\sin\beta - \cos\theta\cos\phi\sin\alpha\sin\beta \right) \\
&+ T\cos\alpha\sin\beta + mV \left( -r\cos\alpha + p\sin\alpha \right)
\end{aligned}
\tag{9}
$$

$$
\begin{aligned}
mV\cos\beta\dot{\alpha} = \ &\bar{q}S \left( C_z \ \cos\alpha - C_x\sin\alpha \right) + mg \left( \sin\theta\sin\alpha + \cos\theta\cos\phi\cos\alpha \right) \\
&- T\sin\alpha + mV \left( q\cos\beta - r\sin\alpha\sin\beta - p\cos\alpha\sin\beta \right)
\end{aligned}
\tag{10}
$$

where Eq. (8) is the $x_w$ component of the vector linear momentum equation, Eq. (9) is its $y_w$ component, and Eq. (10) is its $z_w$ component.

In the above equations, ($m$) is the airplane mass, ($S$) is its wing planform area, ($T$) is the thrust force, ($\theta$) is the Euler's pitch angle, ($g$) is the gravitational acceleration, ($C_x, C_y, C_z$) are nondimensional force coefficients (to be discussed later), and ($\bar{q}$) is the dynamic pressure defined as[164–166]

$$
\bar{q} = \frac{1}{2}\rho V^2
\tag{11}
$$

where ($\rho$) is the air density.

The gravitational acceleration ($g$) is treated here as a constant with the value of 9.81 m/s². Therefore,

$$
g = 9.81\frac{\mathrm{m}}{\mathrm{s}^2}
\tag{12}
$$

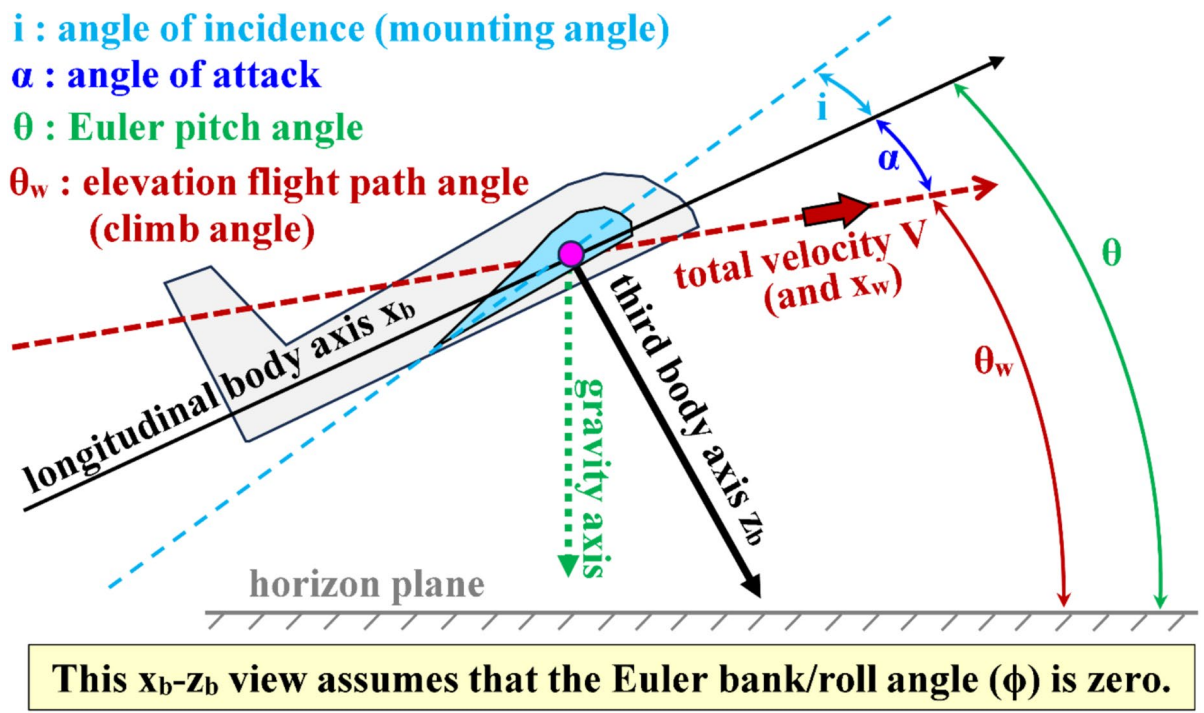

**i : angle of incidence (mounting angle)**
**α : angle of attack**
**θ : Euler pitch angle**
**θw : elevation flight path angle (climb angle)**

This $x_b$-$z_b$ view assumes that the Euler bank/roll angle (φ) is zero.

**Fig. 6.** Illustration of four tilt angles in the midplane of the airplane.

It may be useful to add here that if the linear/translational equations of motion are formulated along the body-fixed axes; then their three components along the longitudinal axis ($x_b$), the lateral body-fixed axis ($y_b$), and the third/bottom body-fixed axis ($z_b$); respectively; become

$$m\dot{u} = X - mg\,\sin\theta + mvr - mwq + T \tag{13}$$

$$m\dot{v} = Y + mg\,\cos\theta\sin\,\phi + mwp - mur \tag{14}$$

$$m\dot{w} = Z + mg\,\cos\theta\cos\,\phi + muq - mvp \tag{15}$$

These are the three components of the following vector linear-momentum equation in the body-fixed axes:

$$m\left(\frac{d}{dt}\left\{\begin{array}{c}u\\v\\w\end{array}\right\} + \left\{\begin{array}{c}p\\q\\r\end{array}\right\} \times \left\{\begin{array}{c}u\\v\\w\end{array}\right\}\right) = \left\{\begin{array}{c}X\\Y\\Z\end{array}\right\} + \left\{\begin{array}{c}mg\,\sin\theta\\mg\,\cos\theta\sin\,\phi\\mg\,\cos\theta\cos\,\phi\end{array}\right\} + \left\{\begin{array}{c}T\\0\\0\end{array}\right\} \tag{16}$$

where ($u, v, w$) are the three components of the velocity vector along the three body-fixed axes ($x_b, y_b, z_b$), respectively. The right-hand side of the above vector equation is the total applied forces on the airplane. The first vector term on the right-hand side represents the aerodynamic forces, the second vector term on the right-hand side represents the weight force, and the third vector term on the right-hand side represents the thrust force.

## Angular-momentum equations

A derived geometric constant ($T_0$) needs to be computed once, and this constant is the determinant of the symmetric inertia tensor. The three diagonal components of this tensor are the moments of inertia (or rectangular moments of inertia) for rotations perpendicular to the three corresponding body axes (centered at the airplane's center of gravity), which are positive numbers for a rigid body; while the off-diagonal elements are the products of inertia, which can be negative, zero, or positive numbers[167–173].

The constant ($T_0$) is expressed mathematically as

$$T_0 = \begin{vmatrix} A & -F & -E \\ -F & B & -D \\ -E & -D & C \end{vmatrix} = ABC - AD^2 - BE^2 - CF^2 - 2DEF \tag{17}$$

The SI unit of each component of the inertia tensor is kg.m$^2$, and thus the SI unit of ($T_0$) is kg$^3$.m$^6$. The six components of the inertia tensor are further explained in Table 1.

For a symmetric airplane, the left (port) half is identical (but reflected) to the right (starboard) half; and in such a case of left–right symmetry, only the product of inertia in the symmetry midplane plane ($x_b - z_b$) is non-zero ($E \neq 0$)[174–176].

Because the rotational equations of motion are expressed in the body-fixed axes and the above inertia terms are formulated also with respect to the body-fixed axes, these inertia terms are considered invariant constants in the rotational equations of motion.

Three auxiliary moments ($T_1, T_2, T_3$) are defined through three algebraic equations as

$$T_1 = (B - C)\,qr + (Eq - Fr)\,p + \left(q^2 - r^2\right)D + L \tag{18}$$

$$T_2 = (C - A)\,rp + (Fr - Dp)\,q + \left(r^2 - p^2\right)E + M \tag{19}$$

$$T_3 = (A - B)\,pq + (Dp - Eq)\,r + \left(p^2 - q^2\right)F + N \tag{20}$$

Finally, the main rotational equations of motion for the airplane about its body axes are

$$T_0\dot{p} = \left(BC - D^2\right)T_1 + (FC + ED)\,T_2 + (FD + EB)\,T_3 \tag{21}$$

$$T_0\dot{q} = \left(AC - E^2\right)T_2 + (AD + EF)\,T_3 + (FC + ED)\,T_1 \tag{22}$$

$$T_0\dot{r} = \left(AB - F^2\right)T_3 + (FD + BE)\,T_1 + (AD + FE)\,T_2 \tag{23}$$

| Inertia symbol | Alternative symbol(s) | Meaning |
|---|---|---|
| A | $I_{xx}$ | Body-referenced moment of inertia about the longitudinal axis ($x_b$) |
| B | $I_{yy}$ | Body-referenced moment of inertia about the lateral axis ($y_b$) |
| C | $I_{zz}$ | Body-referenced moment of inertia about the bottom/third axis ($z_b$) |
| D | $I_{yz}, I_{zy}$ | Body-referenced product of inertia in the plane ($y_b - z_b$) |
| E | $I_{xz}, I_{zx}$ | Body-referenced product of inertia in the plane ($x_b - z_b$) |
| F | $I_{xy}, I_{yx}$ | Body-referenced product of inertia in the plane ($x_b - y_b$) |

**Table 1.** List of the six independent components in the airplane's symmetric inertia tensor.

In the special case of a symmetric airplane, the seven equations presented in the current subsection, namely Eqs. (17–23), can be replaced by only three differential angular-momentum equations that are listed below; which can be derived from Eqs. (17–23) after setting ($D = 0$) and ($F = 0$) and performing some mathematical manipulation[177–185].

$$\dot{p}\left(AC - E^2\right) = \left(BC - E^2 - C^2\right)qr + (A - B + C)\,Epq + CL + EN \tag{24}$$

$$\dot{q}B = Er^2 - Ep^2 + (C - A)\,pr + M \tag{25}$$

$$\dot{r}\left(AC - E^2\right) = \left(A^2 + E^2 - AB\right)pq + (B - A - C)\,Eqr + AN + EL \tag{26}$$

These special simpler equations can be further manipulated and expressed using the alternative symbols for the inertia terms as

$$\dot{p} = \frac{\left(I_{yy}I_{zz} - I_{xz}^2 - I_{zz}^2\right)qr + (I_{xx} - I_{yy} + I_{zz})\,I_{xz}pq + I_{zz}L + I_{xz}N}{I_{xx}I_{zz} - I_{xz}^2} \tag{27}$$

$$\dot{q} = \frac{I_{xz}r^2 - I_{xz}p^2 + (I_{zz} - I_{xx})\,pr + M}{I_{yy}} \tag{28}$$

$$\dot{r} = \frac{\left(I_{xx}^2 + I_{xz}^2 - I_{xx}I_{yy}\right)pq + (I_{yy} - I_{xx} - I_{zz})\,I_{xz}qr + I_{xx}N + I_{xz}L}{I_{xx}I_{zz} - I_{xz}^2} \tag{29}$$

**Inertial velocity**

The equations relating the rates of the ground-referenced inertial coordinates ($\dot{x}_g, \dot{y}_g, \dot{z}_g$) to the velocity magnitude and the spherical flight path angles (azimuth flight path angle $\psi_w$, and elevation flight path angle $\theta_w$) are

$$\dot{x}_g = V\cos\theta_w\cos\psi_w \tag{30}$$

$$\dot{y}_g = V\cos\theta_w\sin\psi_w \tag{31}$$

$$\dot{z}_g = -V\sin\theta_w \tag{32}$$

Figure 7 illustrates the two spherical flight path angles ($\theta_w$ and $\psi_w$), which describe the flying course (direction) of the airplane, treated as a particle, with respect to the inertial origin (the initial trajectory point) using the three spherical coordinates ($V, \psi_w, \theta_w$) as an alternative to the Cartesian inertial velocity components ($\dot{x}_g, \dot{y}_g, \dot{z}_g$).

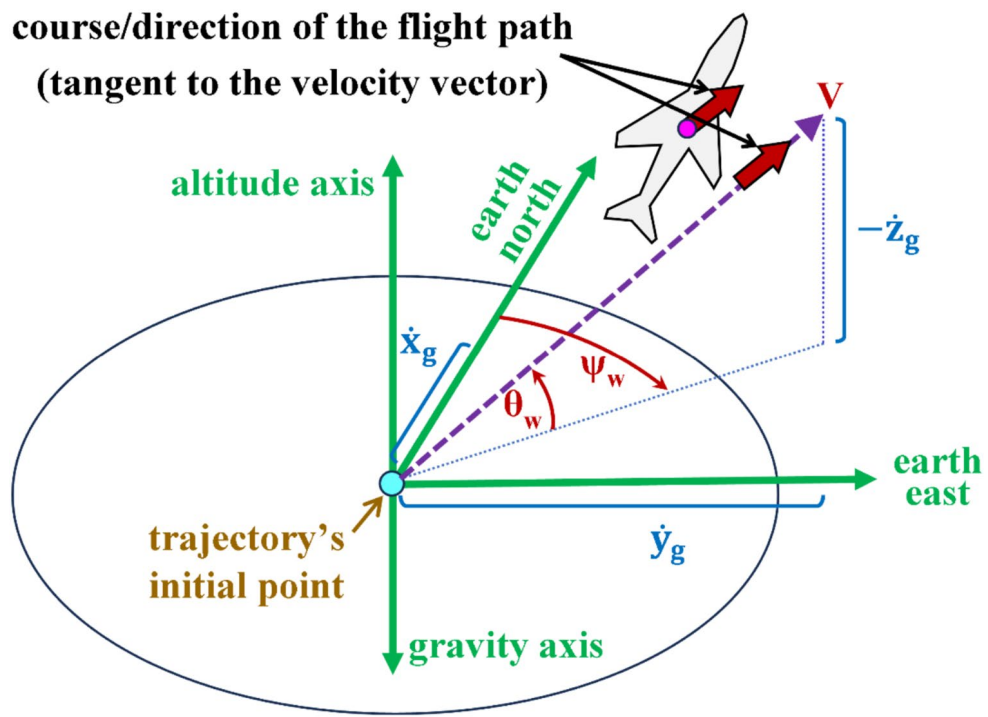

**Fig. 7.** Illustration of the two flight path angles.

## Flight path angles

Two additional algebraic equations relate the two flight path angles $(\psi_w, \theta_w)$ to the three Euler angles $(\phi, \theta, \psi)$ and the two wind axes angles $(\alpha, \beta)$ are provided below[186–191].

$$\cos\theta_w \sin(\psi_w - \psi) = \cos\phi\sin\beta - \sin\phi\sin\alpha\cos\beta \tag{33}$$

$$\sin\theta_w = \sin\theta\cos\alpha\cos\beta - \cos\theta\sin\phi\sin\beta - \cos\theta\cos\phi\sin\alpha\cos\beta \tag{34}$$

From the two above equations; it can be proven that in the condition of equilibrium flight $(\alpha = 0, \beta = 0)$; the Euler yaw angle $(\psi)$, also called "heading angle", becomes equal to the azimuth flight path angle $(\psi_w)$; and the Euler pitch angle $(\theta)$ becomes equal to the elevation flight path angle $(\theta_w)$.

## Three aerodynamic forces

The three body-axes aerodynamic forces acting on the airplane are expressed as

$$X = \overline{q}SC_x \tag{35}$$

$$Y = \overline{q}SC_y \tag{36}$$

$$Z = \overline{q}SC_z \tag{37}$$

In the above equations, $(X)$ is the aerodynamic force along the longitudinal body-fixed axis, and its unit vector exactly coincides with the unit vector of the thrust vector; $(Y)$ is the aerodynamic force along the starboard/right lateral body-fixed axis; and $(Z)$ is the aerodynamic force along the bottom/third body-fixed axis.

## Three moments

The total moment vector is resolved into three components (which we also refer to as "three moments") along the body axes. These moments are expressed in terms of nondimensional moment coefficients $(C_l, C_m, C_n)$ as[192–194]

$$L = \overline{q}SbC_l \tag{38}$$

$$M = \overline{q}ScC_m \tag{39}$$

$$N = \overline{q}SbC_n \tag{40}$$

In the above equations, $(L)$ is the rolling moment about the longitudinal body-fixed axis $(x_b)$, $(M)$ is the pitching moment about the lateral body-fixed axis $(y_b)$, and $(N)$ is the yawing moment about the third/bottom body-fixed axis $(z_b)$. In addition, $(c)$, which is the mean aerodynamic chord (MAC) taken as a characteristic length for nondimensionalizing the pitching moment (longitudinal stability)[195–198].

The mean aerodynamic chord (MAC)[199,200] is defined as follows:

$$\text{cor MAC} = \frac{\int_{\zeta=-b/2}^{b/2} \chi(\zeta)^2 d\zeta}{\int_{\zeta=-b/2}^{b/2} \chi(\zeta) d\zeta} = \frac{1}{S}\int_{\zeta=-b/2}^{b/2} \chi(\zeta)^2 d\zeta = \frac{2}{S}\int_{\zeta=0}^{b/2} \chi(\zeta)^2 d\zeta \tag{41}$$

where $(\chi)$ is the local chord distance as a function of the lateral distance $(\zeta)$, and $(b/2)$ is the semi-span. The mean aerodynamic chord (MAC) is different from the mean geometric chord (MGC), also called standard mean chord (SMC)[201,202], which is defined as

$$\text{MGCorSMC} = \frac{\int_{\zeta=-b/2}^{b/2} \chi(\zeta) d\zeta}{\int_{\zeta=-b/2}^{b/2} d\zeta} = \frac{S}{b} = \frac{1}{b}\int_{\zeta=-b/2}^{b/2} \chi(\zeta) d\zeta = \frac{2}{b}\int_{\zeta=0}^{b/2} \chi(\zeta) d\zeta \tag{42}$$

For most airplanes (with a wing planform resembling a rectangle or a trapezoidal "trapezium"), the values and spanwise location of the standard mean chord (SMC) and the mean aerodynamic chord (MAC) are close to each other; thus, they may be practically treated interchangeably[203]. However, for delta wings, it can be shown that

$$\text{MAC (delta wing)} = \frac{2}{3}C_{root} \tag{43}$$

where $(C_{root})$ is the maximum local chord (at the wing root), while

$$\text{SMC (delta wing)} = \frac{1}{2}C_{root} = \frac{3}{4}\text{MAC (delta wing)} \tag{44}$$

or

$$\text{MAC (delta wing)} = \frac{4}{3}\text{SMC (delta wing)} \tag{45}$$

For a rectangular wing, the local chord is uniform. Thus, the standard mean chord (SMC) and the mean aerodynamic chord (MAC) are exactly equal, or

$$\text{MAC (rectangular wing)} = \text{SMC (rectangular wing)} = \chi \tag{46}$$

It is worth mentioning here that the wing has a nondimensional characteristic attribute, which is the aspect ratio (AR), defined as the square of the span divided by the wing area, or

$$\text{AR} = \frac{b^2}{S} \tag{47}$$

For a rectangular wing, the aspect ratio reduces to the span-to-chord ratio, or

$$\text{AR (rectangular wing)} = \frac{b}{c} \tag{48}$$

The other characteristic length ($b$) pertains to the rolling moment (lateral stability) and the yawing moment (directional stability), and this characteristic nondimensionalization length is the wing span (wingspan)[204–208].

## Aerodynamic and stability coefficients

The aerodynamic lift coefficient ($C_L$), aerodynamic drag coefficient ($C_D$), and aerodynamic side-force coefficient ($C_C$) are wind-axes nondimensional quantities from which the body-axes aerodynamic coefficients ($C_x, C_y, C_z$) can be obtained[209–216].

The lift coefficient is modeled here as being directly related to the conventional angle of attack ($\widetilde{\alpha}$) as

$$C_L(\widetilde{\alpha}) = C_{L0} + C_{L\alpha}\widetilde{\alpha} \tag{49}$$

where ($C_{L0}$) is the lift coefficient at zero conventional angle of attack, and ($C_{L\alpha}$) is the gain in the lift coefficient per unit increase in the angle of attack (when expressed in radians), and both values are treated as constant parameters.

As mentioned earlier in "Linear-momentum equations and equilibrium" section, the term "angle of attack" ($\alpha$) used in the proposed flight mechanics simulation modeling here is the change (either positive or negative) in the conventional angle of attack from its equilibrium value ($\widetilde{\alpha}_{equb}$), which is assumed to be equal to the angle of incidence for the wing ($i$). Therefore, the lift coefficient ($C_L$) is related to the modeling/simulation angle of attack ($\alpha$) as

$$C_L(\alpha) = C_{L0} + C_{L\alpha}(\alpha + \widetilde{\alpha}_{equb}) \tag{50}$$

Regardless of the lift coefficient being expressed as a function of the conventional angle of attack ($\widetilde{\alpha}$) as in Eq. (49), or being expressed as a function of the modeling/simulation angle of attack ($\alpha$) as in Eq. (50), it has the same slope ($C_{L\alpha}$). At any given value of ($C_L$), the angular difference ($\widetilde{\alpha} - \alpha$) is equal to the equilibrium value of the conventional angle of attack ($\widetilde{\alpha}_{equb}$); and it is computed from the force balance between the weight of the airplane ($mg$) and the equilibrium lift force ($\overline{q}SC_{L,equb}$), where ($C_{L,equb}$) is the lift coefficient at equilibrium flight. Therefore,

$$\frac{mg}{\overline{q}S} = C_{L,equb} \tag{51}$$

Combining the above defining equation for the equilibrium lift coefficient with Eq. (49) that relates the lift coefficient ($C_L$) linearly with the conventional angle of attack ($\widetilde{\alpha}$) gives

$$\frac{mg}{\overline{q}S} = C_{L0} + C_{L\alpha}\widetilde{\alpha}_{equb} \tag{52}$$

This leads to an explicit expression for the conventional equilibrium angle of attack ($\widetilde{\alpha}_{equb}$) as

$$\widetilde{\alpha}_{equb} = \frac{\frac{mg}{\overline{q}S} - C_{L0}}{C_{L\alpha}} \tag{53}$$

This conventional equilibrium angle of attack ($\widetilde{\alpha}_{equb}$) is approximately constant if the dynamic pressure ($\overline{q}$) changes only slightly, and this implies that the air density and the flight speed are nearly unchanged, and this is a valid assumption in steady-level flight.

Figure 8 illustrates the linear dependence of the lift coefficient ($C_L$) on the modeling/simulation angle of attack ($\alpha$), which is the default angle of attack in the current work. The figure also illustrates the linear dependence of the lift coefficient ($C_L$) on the conventional angle of attack ($\widetilde{\alpha}$)[217–220]. This linear dependence is appropriate as long as the airplane is away from aerodynamic stall conditions[221–227].

Through the drag polar relationship, the drag coefficient depends on the lift coefficient as[228–230]

$$C_D = C_{D0} + K_{CD}C_L^2 \tag{54}$$

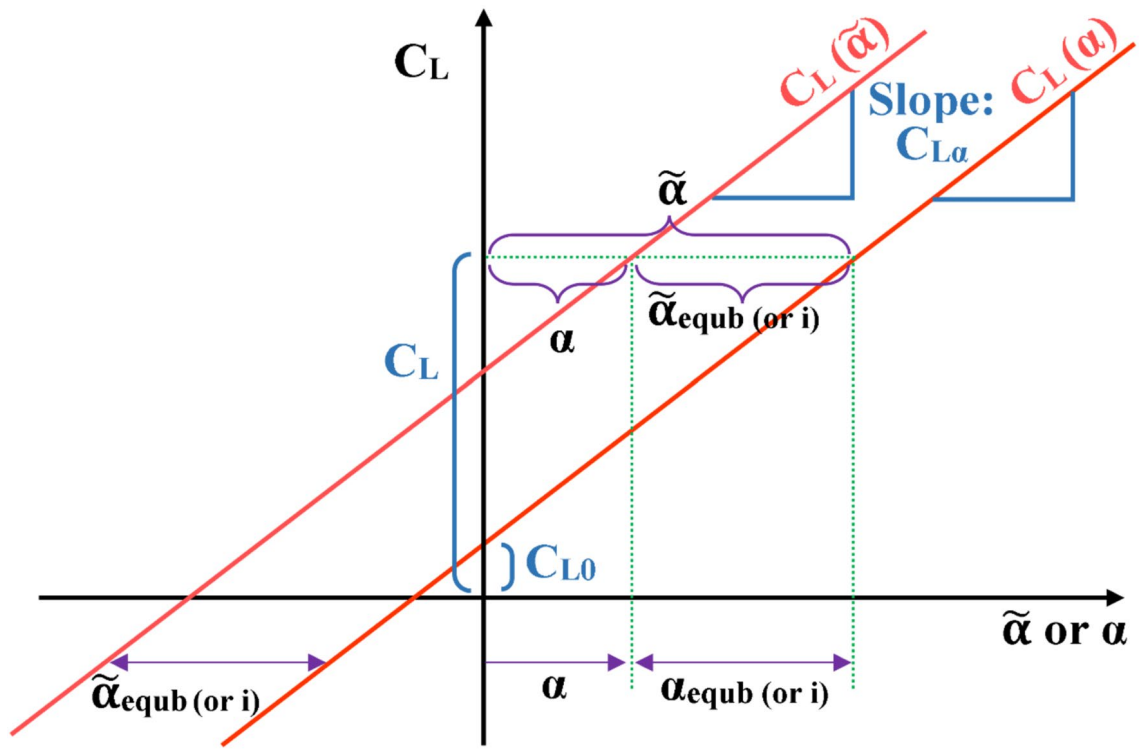

**Fig. 8.** Illustration of the lift coefficient profile.

where $(C_{D0})$ and $(K_{CD})$ are additional aerodynamic constants.

The side-force coefficient depends on the sideslip angle as

$$C_C = C_{C\beta}\beta \tag{55}$$

After knowing the wind-axes aerodynamic coefficients $(C_L, C_D, C_C)$, the body-axes aerodynamic coefficients $(C_x, C_y, C_z)$ can be obtained using straightforward trigonometric projections as

$$C_x = -C_D\cos\alpha\cos\beta - C_C\cos\alpha\sin\beta + C_L\sin\alpha \tag{56}$$

$$C_y = -C_D\sin\beta + C_C\cos\beta \tag{57}$$

$$C_z = -C_D\sin\alpha\cos\beta - C_C\sin\alpha\sin\beta - C_L\cos\alpha \tag{58}$$

The body-axes total-moment coefficients $(C_l, C_m, C_n)$ are modeled as[231]

$$C_l = C_{l\beta}\beta + C_{lp}p(b/V) + C_{lr}r(b/V) + C_{l\delta l}\delta_l + C_{l\delta n}\delta_n \tag{59}$$

$$C_m = C_{m0} + C_{m\alpha}\alpha + C_{mq}q\,(c/V) + C_{m\delta m}\delta_m \tag{60}$$

$$C_n = C_{n\beta}\beta + C_{np}p(b/V) + C_{nr}r(b/V) + C_{n\delta l}\delta_l + C_{n\delta n}\delta_n \tag{61}$$

where $(C_{lp}, C_{lr}, C_{m0}, C_{np}, C_{nr})$ are nondimensional constant parameters, while $(C_{l\beta}, C_{l\delta l}, C_{l\delta n}, C_{m\alpha}, C_{m\delta m}, C_{n\beta}, C_{n\delta l}, C_{n\delta n})$ are dimensional constant parameters having the unit of 1/rad, $(\delta_l)$ is the ailerons' deflection angle in radians, $(\delta_m)$ is the elevators' deflection angle in radians, and $(\delta_n)$ is the rudder's deflection angle in radians.

### Air density and speed of sound

The variation of the air density with the altitude $(h)$ is governed here by the International Standard Atmosphere (ISA) model. Up to an altitude of about 11,000 m above mean sea level, the troposphere layer of the air is present, in which the air temperature decreases linearly with altitude, while the density declines nonlinearly with altitude at a faster rate according to[232–239].

$$\rho_{troposphere} = 1.225\left(1 - \frac{\lambda}{288.15}h\right)^{g/(R\lambda)-1} = 1.225\left(1 - 2.2558 \times 10^{-5}h\right)^{4.2586} \tag{62}$$

where $(\lambda)$ is the lapse rate magnitude, taken as 0.0065 K/m; and $(R)$ is the ideal gas constant for air, which is the universal molar gas constant divided by the molecular weight, and this gas constant for air is 287 J/kg.K[240–243]. In the previous equation, the constant 1.225 is the standard sea level air density in kg/m$^3$ ($\rho_{h=0}$); and the constant 288.15 is the standard sea level absolute temperature in kelvins ($\Theta_{h=0}$), which corresponds to 15 °C[244–248]. The standard sea level atmospheric pressure in the ISA model is 101,325 Pa (101.325 kPa, 1.01325 bar, or 760 mmHg)[249–252].

| Geometric altitude (m) | Geopotential altitude (m) | Absolute difference (m) | Percentage deviation |
|---|---|---|---|
| 5,000.000 | 4,996.079 | 3.921 | 0.0785% |
| 5,003.927 | 5,000.000 | 3.927 | 0.0785% |
| 10,000.000 | 9,984.328 | 15.672 | 0.1568% |
| 10,015.721 | 10,000.000 | 15.721 | 0.1571% |

**Table 2.** Examples of the geometric and geopotential altitudes within the troposphere layer.

| | |
|---|---|
| Air density computed using an altitude of 4996 m ($\rho_{4996}$) | 0.736191 kg/m$^3$ |
| Air density computed using an altitude of 5000 m ($\rho_{5000}$) | 0.735872 kg/m$^3$ |
| Air density computed using an altitude of 5004 m ($\rho_{5004}$) | 0.735553 kg/m$^3$ |
| $\frac{\rho_{4996} - \rho_{5000}}{\rho_{5000}} \times 100\%$ | 0.0433% |
| $\frac{\rho_{5004} - \rho_{5000}}{\rho_{5000}} \times 100\%$ | −0.0433% |

**Table 3.** Quantification of the error in air density calculation at altitudes near 5,000 m.

| | |
|---|---|
| Air density computed using an altitude of 9984 m ($\rho_{9984}$) | 0.413234 kg/m$^3$ |
| Air density computed using an altitude of 10,000 m ($\rho_{10000}$) | 0.412415 kg/m$^3$ |
| Air density computed using an altitude of 10,016 m ($\rho_{10016}$) | 0.411597 kg/m$^3$ |
| $\frac{\rho_{9984} - \rho_{10000}}{\rho_{10000}} \times 100\%$ | 0.1986% |
| $\frac{\rho_{10016} - \rho_{10000}}{\rho_{10000}} \times 100\%$ | −0.1983% |

**Table 4.** Quantification of the error in air density calculation at altitudes near 10,000 m.

Although the ISA model is based on geopotential altitudes for estimating the air density, we here utilize the geometric (true or orthometric) altitude[253–265]. This simplification is aligned with our treatment of the gravitational acceleration as a universal constant. We assessed the difference in the two types of altitudes, and we found that the difference in the troposphere layer of interest here is small, as shown in Table 2.

The percentage deviations in the above table are very small, much less than 1%. These percentage deviations were computed as[266]

$$\%\text{deviation} = 2\frac{\left|\text{geometric altitude } - \text{ geopotential altitude}\right|}{\text{geometric altitude } + \text{ geopotential altitude}} \times 100\% \tag{63}$$

The relationship between the geometric altitude ($h$) and geopotential altitude (denoted by $H$) is

$$H = \frac{R_E h}{R_E + h} \tag{64}$$

$$h = \frac{R_E H}{R_E - h} \tag{65}$$

where ($R_E$) is the mean radius of the earth, and it is taken here as 6,371,000 m [267–271].

In Table 3, we further assess the influence of using a geometric altitude rather than a geopotential altitude when using Eq. (62) for computing the air density. At a geometric altitude of 5,000 m, for example, the air density should be computed strictly speaking at the slightly-lower geopotential altitude of approximately 4,996 m. Similarly, the air density computed at a geopotential altitude of 5,000 m strictly speaking corresponds to the slightly higher geometric altitude of approximately 5,004 m. However, the values in the table show that these three densities are nearly the same, and the error incurred by the simplifying assumption of using the geometric altitude in lieu of the geopotential altitude leads to a marginal error in the air density that is below 0.05% at altitudes near 5,000 m; corresponding approximately to the middle of the troposphere layer of the atmosphere.

Similarly, Table 4 assesses the expected error in the air density but at higher altitudes near 10,000 m; located close to the upper end of the troposphere layer of the atmosphere. Although the error grows (nonlinearly) as the altitude increases, it remains very small, below 0.2%. Regarding the three altitudes listed in this table (9,984 m; 10,000 m; and 10,016 m); the geopotential altitude corresponding to a geometric altitude of 10,000 m is approximately 10,016 m; while the geometric altitude corresponding to a geopotential altitude of 10,000 m is approximately 9,984 m.

Many commercial transport airplanes fly at cruising altitudes below 11 km, thus the troposphere formula is adequate for them[272–280]. For higher altitudes, as in some military aircraft, another expression for the air density should be used, which corresponds to the tropopause layer of the atmosphere; it is an interspheric layer lying between the lower troposphere layer and the upper stratosphere layer, extending approximately between the altitudes 11,000 m and 20,000 m[281–286]. In the tropopause layer of the atmosphere, the temperature is treated as constant (with a value of –56.50 °C or 216.65 K), making the tropopause an isothermal layer. Therefore, the decline of the air density with the altitude within the tropopause layer follows a different profile than the one described earlier for the non-isothermal troposphere layer. This decline is described as

$$\rho_{tropopause} = 0.3636309 \exp\left(-\frac{g}{216.65R}(h - 11,000)\right)$$
$$= 0.3636309 e^{-1.5777145 \times 10^{-4}(h - 11,000)} \tag{66}$$

where the value (11,000) is the altitude (in meters) at the bottom edge of the tropopause layer, which is also the upper edge of the troposphere layer; the value (0.3636309) is the air density at this bottom edge of the troposphere layer ($\rho_{h=11\,km}$) in kg/m³, as computed by the tropospheric density equation; and the value (216.65) is the absolute temperature (in kelvins) of air within the tropopause layer ($\Theta_{h=11-20\,km}$).

It is worth mentioning that the decline of the air density in the non-isothermal troposphere layer is slower than its decline in the isothermal tropopause layer, because the drop of temperature in the non-isothermal troposphere layer has a compressive effect on air, causing its density to tend to increase; however, the drop in the pressure due to the diminishing weight of the above air column causes a stronger expansive effect of decreasing the air density; and resultantly the air density declines with the altitude[287–291]. In the isothermal tropopause layer, the compressive temperature effect is eliminated, thereby magnifying the expansive pressure effect, and thus the air density declines faster with the altitude. To demonstrate this faster rate of density decline in the isothermal tropopause layer, applying Eq. (66) describing the tropopause's air density at an altitude of zero ($h= 0$), which belongs to the troposphere layer – outside the tropopause layer, gives an extrapolated benchmarking value for the sea-level air density of 2.0624 kg/m³, which clearly exceeds the true sea level value of 1.225 kg/m³[292–294].

Although the speed of sound ($a$) in air is not an essential variable in the presented flight mechanics model, it is still an important property in aeronautics, because the ratio between the airplane speed ($V$) and the speed of sound is the nondimensional Mach number ($M$) that serves as a criterion for determining the regime of flight as well as the possible occurrence of special phenomena such as shock waves[295–299]. The Mach number is defined as

$$M = \frac{V}{a} \tag{67}$$

The flight regime with $M < 1$ is subsonic, the flight regime with $M \approx 1$ is transonic or sonic, the flight regime with $1 > M \geq 5$ is supersonic, while the condition $M > 5$ corresponds to the hypersonic flight regime[300–304].

The speed of sound for air (as an ideal gas) depends on its absolute temperature ($\Theta$) and on its specific heat ratio (the ratio of specific heat capacities, or the adiabatic index, $\gamma$) as[305–307]

$$a = \sqrt{\gamma R \Theta} \tag{68}$$

In the troposphere layer of the atmosphere (the non-isothermal layer adjacent to the ground), the absolute air temperature is computed according to the lapse rate ($\lambda = 0.0065$ K/m) as

$$\Theta_{troposphere} = 288.15 - \lambda h \tag{69}$$

In the tropopause layer of the atmosphere (the isothermal layer next to the troposphere layer), the absolute air temperature is assumed to have a constant value of 216.65 K. Thus,

$$\Theta_{tropopause} = 216.65 \text{K} \tag{70}$$

The specific heat ratio is treated here as a constant with a value of 1.4, which is commonly assigned to ambient air[308–310].

$$\gamma = 1.4 \tag{71}$$

## Summary of equations, variables, and constants
In the current section, we provide a summary of the overall six-degree-of-freedom (6-DOF) fluid mechanics problem in terms of the mathematical structure.

### Summary of equations
The current fluid mechanics problem is described by a nonlinear differential–algebraic equations (DAE) system, consisting of 35 equations that are either ordinary differential equations (ODE) or algebraic equations. These 35 scalar equations can be grouped into nine categories as shown in Table 5. It should be noted that these groups are not mathematically decoupled; the grouping is based on the scope of use for the equations.

| Equations group | Equations count |
|---|---|
| Body-fixed axes angular velocity components | 3 |
| Wind-axes linear-momentum equations (and the dynamic pressure) | 4 |
| Body-axes angular-momentum equations (including the auxiliary moments) | 6 |
| Inertial velocity components | 3 |
| Flight path angles | 2 |
| Body-fixed axes aerodynamic forces | 3 |
| Body-fixed axes total moments | 3 |
| Aerodynamic and stability (moment) coefficients | 9 |
| Air density (and the flight altitude) | 2 |
| Total | 35 |

**Table 5.** Summary of the 35 flight mechanics equations in the current study.

| Variables group | Variables type (in InvSim) | Variables symbols | Variables count |
|---|---|---|---|
| Inertial coordinates and **roll** Euler angle | input | $x_g, y_g, z_g, \phi$ | 4 |
| Pitch and yaw Euler angles | intermediate | $\theta, \psi$ | 2 |
| Body-axes angular velocity components | intermediate | $p, q, r$ | 3 |
| Wind-axes coordinates for the linear velocity | intermediate | $V, \alpha, \beta$ | 3 |
| Spherical angular coordinates (flight path angles) for the airplane's inertial location | intermediate | $\theta_w, \psi_w$ | 2 |
| Body-axes aerodynamic forces (and dynamic pressure) | intermediate | $X, Y, Z, \bar{q}$ | 4 |
| Body-axes total moments (and auxiliary moments) | intermediate | $T_1, T_2, T_3, L, M, N$ | 6 |
| Aerodynamic and stability coefficients | intermediate | $C_L, C_D, C_C, C_x, C_y, C_z, C_l, C_m, C_n$ | 9 |
| Air density and flight altitude | intermediate | $\rho, h$ | 2 |
| Flight controls | output | $T, \delta_l, \delta_m, \delta_n$ | 4 |
| Total | | | 39 |

**Table 6.** Summary of the 39 flight mechanics variables in the current study.

## Summary of variables

The aforementioned 35 flight mechanics equations have 39 independent variables, which are listed in Table 6 as categorized groups. It should be noted that constant parameters that appear in the differential–algebraic equations (such as the wing planform area and the airplane mass) are not included among the flight variables, because these parameters remain invariant during the entire flight trajectory. In addition, the time derivative of a flight variable is not considered an additional separate variable.

The difference between the number of equations and the number of variables is four, and this is the number of input constraints that should be specified in order to be able to integrate the flight mechanics system and obtain a unique solution. The four constraints in the case of our inverse simulation (InvSim) flight mechanics model are the temporal profiles of the three inertial coordinates ($x_g, y_g, z_g$) and the temporal profile of the roll angle ($\phi$). The output variables to be obtained by the InvSim flight mechanics solver are the discrete profiles of four airplane controls, which are the thrust and the three deflection angles of the control surfaces ($T, \delta_l, \delta_m, \delta_n$). The remaining 31 variables (such as the angle of attack $\alpha$, and the Euler pitch angle $\theta$) are intermediate quantities that can evolve over time during the flight maneuver in response to changes in other related variables.

## Summary of constants

In addition to the 39 flight variables (that generally vary during the flight maneuver), various constant parameters need to be defined once, and these values remain unchanged during the entire flight simulation. The total number of constants in the proposed InvSim model is 30, which are classified as 29 parameters for describing the airplane and its performance, and one parameter related to the trajectory (the initial altitude). These 30 constant parameters are summarized in Table 7, which are organized as related groups.

It should be noted that eight physical universal constants are not counted among the 30 parameters, because these are not customizable quantities. These universal constants (for the proposed InvSim algorithm) are

1. Gravitational acceleration ($g = 9.81$ m/s$^2$)
2. Tropospheric lapse rate magnitude ($\lambda = 0.0065$ K/m)
3. Ideal gas constant for air ($R = 287$ J/kg.K)
4. Standard sea-level air density ($\rho_{h=0} = 1.225$ kg/m$^3$)
5. Standard sea-level air absolute temperature ($\Theta_{h=0} = 288.15$ K)
6. Standard altitude of the troposphere-tropopause transition (11,000 m)
7. Standard air density at the troposphere-tropopause transition ($\rho_{h=11km} = 0.3636309$ kg/m$^3$)
8. Standard air absolute temperature within the tropopause layer ($\Theta_{h=11-20km} = 216.65$ K)

| Parameters group | Parameters symbols | Parameters count |
|---|---|---|
| Airplane mass | $m$ | 1 |
| Wing planform (projected) area | $S$ | 1 |
| Mean aerodynamic chord (MAC) | $c$ | 1 |
| Wing span | $b$ | 1 |
| Mass moments and products of inertia about body axes | $A, B, C, D, E, F$ | 6 |
| Aerodynamic-force constants | $C_{L0}, C_{L\alpha}, C_{D0}, K_{CD}, C_{C\beta}$ | 5 |
| Longitudinal stability derivatives | $C_{m0}, C_{m\alpha}, C_{mq}, C_{m\delta m}$ | 4 |
| Lateral stability derivatives | $C_{l\beta}, C_{lp}, C_{lr}, C_{l\delta l}, C_{l\delta n}$ | 5 |
| Directional stability derivatives | $C_{n\beta}, C_{np}, C_{nr}, C_{n\delta l}, C_{n\delta n}$ | 5 |
| Initial altitude | $h_{ini}$ | 1 |
| Total | | 30 |

**Table 7.** Summary of the 30 constant parameters for the presented inverse simulation model.

While it is possible to upgrade the model by treating the gravitational acceleration as a function of altitude (in this case, the number of flight variables increases from 39 to 40); the gain from this upgrade is not justified. This change largely increases the mathematical complexity of the model, where the time derivative of the gravitational acceleration appears in the mathematical expressions, while such a derivative is practically zero. To demonstrate this (and to justify treating the gravitational acceleration as a universal constant), we first point out that the altitude-dependent gravitational acceleration ($\widetilde{g}$) declines with the geometric altitude according to the following quadratic relationship:

$$\frac{\widetilde{g}(h)}{g} = \left(\frac{R_E}{R_E + h}\right)^2 \tag{72}$$

At a geometric altitude of 11,000 m (which is the upper limit of the troposphere layer), the altitude-dependent gravitational acceleration is 99.656% of its approximated constant value (at sea level); and at a higher geometric altitude of 20,000 m (which is the upper limit of the tropopause layer), the altitude-dependent gravitational acceleration drops further to 99.375% its approximated constant value. Thus, the relative drops in the gravitational acceleration at 11,000 m and 20,000 m geometric altitudes are only 0.344% and 0.625%; respectively. As dimensional drops in the altitude-dependent gravitational acceleration from an assigned sea-level value of 9.81 m/s²; the respective drops at 11,000 m and 20,000 m geometric altitudes of 9.81 m/s² ($g - \widetilde{g}(11,000m) = 9.81$ m/s² $- 9.776212$ m/s²) and 0.0061303 m/s² ($g - \widetilde{g}(20,000m) = 9.81 - 9.748697$ m/s²); respectively. At a rate-of-climb (RoC) of 2000 ft/min (33.33 ft/s, 10.16 m/s, or 36.58 km/h), which is typical for a commercial jet airplane, an altitude of 11,000 m (36,089 ft) can be reached after a continuous climb for about 1083 s (about 18 min); and this means that the average rate of change of the altitude-dependent gravitational acceleration is approximately $3.12 \times 10^{-6}$ m/s³, which is nearly zero[311–318]. Although high-performance fighter airplanes may achieve much higher climb rates and descent rates, such as 20,000 ft/min (10 times the typical rates for commercial jet airplanes), the time rate of change in the altitude-dependent gravitational acceleration in such cases remains negligible[319–325].

9. Specific heat ratio for air ($\gamma = 1.4$)

Although the specific heat ratio for air (as an ideal non-monatomic gas) is actually a function of its temperature (while it is temperature-independent or monatomic gases like helium and argon[326,327]), and this air temperature is a function of the altitude within the troposphere layer; treating the specific heat ratio as a universal constant as adopted here is appropriate given the narrow range of temperatures for atmospheric air during an airplane flight[328–330]. The specific heat ratio is the ratio of the specific heat capacity at constant pressure ($C_p$) to the specific heat capacity at constant volume ($C_v$)[331,332]. In this case of a constant specific heat ratio, air is assumed to be a "calorically perfect" gas, which is a special simplified case of ideal gases[333]. For air, the specific heat ratio is 1.4022 at 288.15 K (15.00 °C) and it increases to 1.4027 at 216.65 K (−56.50 °C); while it drops to 1.4015 at 323.15 K (50.00 °C); and these values show the weak deviation from the nominal used value of 1.4[334].

## InvSim customized equations of motion

In this part, we present a variant of the general fluid mechanics equations of motion, which are adapted for use in an inverse simulation (InvSim) mode. These customized equations are presented as nine groups in the next nine subsections, with a different order than the one presented earlier for the general fluid mechanics equations (not customized for the inverse mode) in "General equations of motion" section; this change in the order of presentation facilitates explaining the rationale behind the need for additional derivative expressions beyond those appearing in the general flight mechanics equations.

As a summary before starting the detailed formulation, the three unknown control deflection angles $(\delta_l, \delta_m, \delta_n)$ are computed from three algebraic equations, while the unknown thrust force $(T)$ is computed among eight flight variables through a classical fourth-order Runge–Kutta method (RK4) for integrating a system of nonlinear ordinary differential equations (ODE)[335–340].

## InvSim aerodynamic and stability coefficients

The formulas presented earlier in "Aerodynamic and stability coefficients" section for the aerodynamic lift coefficient $(C_L)$, the aerodynamic drag coefficient $(C_D)$, the side-force coefficient $(C_C)$, and the three body-axes aerodynamic coefficients $(C_x, C_y, C_z)$ remain in use in the transformed InvSim mode of the flight mechanics equations of motion. For completeness, these expressions are repeated here (while being assigned their same original equation numbers).

$$C_L = C_{L0} + C_{L\alpha}\alpha \tag{49}$$

$$C_D = C_{D0} + K_{CD}C_L^2 \tag{54}$$

$$C_C = C_{C\beta}\beta \tag{55}$$

$$C_x = -C_D\cos\alpha\cos\beta - C_C\cos\alpha\sin\beta + C_L\sin\alpha \tag{56}$$

$$C_y = -C_D\sin\beta + C_C\cos\beta \tag{57}$$

$$C_z = -C_D\sin\alpha\cos\beta - C_C\sin\alpha\sin\beta - C_L\cos\alpha \tag{58}$$

However, the original expressions for the body-axes total-moment coefficients $(C_l, C_m, C_n)$ are restructured into three explicit expressions to obtain the three control surface deflection angles. The original expressions for the body-axes total-moment coefficients $(C_l, C_m, C_n)$ are repeated below to facilitate the derivation of the restructured ones.

$$C_l = C_{l\beta}\beta + C_{lp}p(b/V) + C_{lr}r(b/V) + C_{l\delta l}\delta_l + C_{l\delta n}\delta_n \tag{59}$$

$$C_m = C_{m0} + C_{m\alpha}\alpha + C_{mq}q\,(c/V) + C_{m\delta m}\delta_m \tag{60}$$

$$C_n = C_{n\beta}\beta + C_{np}p(b/V) + C_{nr}r(b/V) + C_{n\delta l}\delta_l + C_{n\delta n}\delta_n \tag{61}$$

From Eq. (60), the restructured explicit expression for the necessary control surface deflection angle for the elevators $(\delta_m)$ can be obtained as

$$\delta_m = \frac{C_m - C_{m0} - C_{m\alpha}\alpha - C_{mq}q\frac{c}{V}}{C_{m\delta m}} \tag{73}$$

Solving Eqs. (59 and 61) simultaneously for the necessary control surface deflection angle of the ailerons $(\delta_l)$ and the rudder $(\delta_n)$ gives

$$\delta_l = \frac{C_{n\delta n}\left(C_l - C_{l\beta}\beta - C_{lp}p\frac{b}{V} - C_{lr}r\frac{b}{V}\right) - C_{l\delta n}\left(C_n - C_{n\beta}\beta - C_{np}p\frac{b}{V} - C_{nr}r\frac{b}{V}\right)}{C_{l\delta l}C_{n\delta n} - C_{l\delta n}C_{n\delta l}} \tag{74}$$

$$\delta_n = \frac{C_{l\delta l}\left(C_n - C_{n\beta}\beta - C_{np}p\frac{b}{V} - C_{nr}r\frac{b}{V}\right) - C_{n\delta l}\left(C_l - C_{l\beta}\beta - C_{lp}p\frac{b}{V} - C_{lr}r\frac{b}{V}\right)}{C_{l\delta l}C_{n\delta n} - C_{l\delta n}C_{n\delta l}} \tag{75}$$

The coupling between the rolling and yawing moments is noticeable from Eqs. (59 and 61), and these two moments are together decoupled from the pitching moment as indicated by Eq. (60); this behavior is known for fixed-wing airplanes[341–346].

## InvSim three moments

In "Three moments" section of the original general (not customized for the inverse mode) flight mechanics formulation, the total moment vector was resolved into three components (three moments; $L, M, N$) along the body axes $(x_b, y_b, z_b)$; and these body-referenced moments were expressed in terms of three nondimensional moment coefficients $(C_l, C_m, C_n)$, respectively. However, in the previous "InvSim aerodynamic and stability coefficients" section, we showed that the proposed InvSim flight mechanics formulation requires the values of these nondimensional moment coefficients to algebraically obtain the corresponding control surface deflection angles $(\delta_l, \delta_m, \delta_n)$.

Therefore, the original expressions for $(L, M, N)$ are inverted here to be explicit expressions for obtaining $(C_l, C_m, C_n)$, as follows:

$$C_l = \frac{L}{\bar{q}Sb} \tag{76}$$

$$C_m = \frac{M}{\bar{q}Sc} \tag{77}$$

$$C_n = \frac{N}{\bar{q} S b} \tag{78}$$

### InvSim angular-momentum equations

In the part of the original general flight mechanics formulation covered earlier in "Three moments" section, algebraic expressions for three auxiliary moments $(T_1, T_2, T_3)$ were given as functions of the body-referenced angular velocity components $(p, q, r)$ and body-referenced total moments $(L, M, N)$; with the body-axes inertia components $(A, B, C, D, E, F)$ being constant geometric parameters. These expressions are repeated below.

$$T_1 = (B - C)\, qr + (Eq - Fr)\, p + \left(q^2 - r^2\right) D + L \tag{18}$$

$$T_2 = (C - A)\, rp + (Fr - Dp)\, q + \left(r^2 - p^2\right) E + M \tag{19}$$

$$T_3 = (A - B)\, pq + (Dp - Eq)\, r + \left(p^2 - q^2\right) F + N \tag{20}$$

Also, the original general flight mechanics formulation in "Three moments" section included three main rotational equations of motion for the airplane about its body axes, which are repeated below.

$$T_0 \dot{p} = \left(BC - D^2\right) T_1 + (FC + ED)\, T_2 + (FD + EB)\, T_3 \tag{21}$$

$$T_0 \dot{q} = \left(AC - E^2\right) T_2 + (AD + EF)\, T_3 + (FC + ED)\, T_1 \tag{22}$$

$$T_0 \dot{r} = \left(AB - F^2\right) T_3 + (FD + BE)\, T_1 + (AD + FE)\, T_2 \tag{23}$$

As explained before, $(T_0)$ is the determinant of the inertia tensor.

However, according to the discussion given in the previous subsection, the body-referenced total moments $(L, M, N)$ are needed in order to obtain the corresponding nondimensional moment coefficients $(C_l, C_m, C_n)$. These body-referenced total moments are obtained by solving simultaneously the above system of three equations, Eqs. (21–23), for the auxiliary moments $(T_1, T_2, T_3)$ as three explicit expressions whose right-hand side include the body-referenced angular accelerations $(\dot{p}, \dot{q}, \dot{r})$. The resultant explicit symbolic expressions for $(T_1)$ and $(T_2)$ are very large and thus are not shown here; but we designate them by the placeholder symbols $(T_{1,\mathcal{F}})$ and $(T_{2,\mathcal{F}})$, respectively (the subscript "$\mathcal{F}$" here indicates a function that can be simply evaluated to yield a left-hand side variable if all right-hand side elements are known). However, the explicit symbolic expressions for $(T_3)$ is simple enough to be shown here, and it is given below and we designate it by the symbol $(T_{3,\mathcal{F}})$ to distinguish it from the original explicit expression in Eq. (20) that also has $(T_3)$ in its left-hand side as a single quantity, but that original expression is not used in the proposed InvSim algorithm presented here, because it is replaced by the derived expression below for $(T_{3,\mathcal{F}})$ in Eq. (79).

$$T_{3,\mathcal{F}} : T_3 = T_0 \frac{\dot{q}D - \dot{r}C + \dot{p}E}{AD^2 + 2DEF - ACB + F^2 C - E^2 B} \tag{79}$$

After the values of the auxiliary moments become known at a given time station, the body-referenced total moments $(L, M, N)$ can be computed using an adapted version of the equations that defined the auxiliary moments $(T_1, T_2, T_3)$; namely Eqs. (18, 19, 20), respectively. These adapted equations are suitable for computing the numerical values of $(L, M, N)$, and they have the following form:

$$L = T_1 - (B - C)\, qr - (Eq - Fr)\, p - \left(q^2 - r^2\right) D \tag{80}$$

$$M = T_2 - (C - A)\, rp - (Fr - Dp)\, q - \left(r^2 - p^2\right) E \tag{81}$$

$$N = T_3 - (A - B)\, pq - (Dp - Eq)\, r - \left(p^2 - q^2\right) F \tag{82}$$

### InvSim angular velocity vector in body axes

The expressions relating the body-referenced roll rate $(p)$, body-referenced pitch rate $(q)$, and body-referenced yaw rate $(r)$ to the Euler rates and Euler angles remain the same. These expressions were presented in "Angular velocity vector in body axes" section through Eqs. (2–4), which are repeated below.

$$p = \dot{\phi} - \sin\theta\, \dot{\psi} \tag{2}$$

$$q = \cos\phi\, \dot{\theta} + \cos\theta\sin\phi\, \dot{\psi} \tag{3}$$

$$r = \cos\theta\cos\phi\, \dot{\psi} - \sin\phi\, \dot{\theta} \tag{4}$$

The above algebraic expressions for the body-referenced angular velocities $(p, q, r)$ are used to find initial values $(p_{ini}, q_{ini}, r_{ini})$ for them at the beginning of the maneuver's numerical simulation, at the time $(t) = 0$.

According to the InvSim customized expressions $(T_{1,\mathcal{F}}, T_{2,\mathcal{F}}, T_{3,\mathcal{F}})$ discussed earlier in the previous "InvSim angular-momentum equations" section for the auxiliary moments $(T_1, T_2, T_3)$, it is required to know the values of the derivatives of the body-referenced angular velocities $(\dot{p}, \dot{q}, \dot{r})$ in order to be able to compute the dependent values of the auxiliary moments $(T_1, T_2, T_3)$. Expressions for these three body-referenced angular accelerations $(\dot{p}, \dot{q}, \dot{r})$ are obtained symbolically by direct differentiation of the expressions for $(p, q, r)$ as presented in Eqs. (2–4). The resultant expressions for the body-referenced angular accelerations are

$$\dot{p} = \ddot{\phi} - \cos\theta\dot{\psi}\dot{\theta} - \sin\theta\ddot{\psi} \tag{83}$$

$$\dot{q} = -\sin\phi\dot{\theta}\dot{\phi} + \cos\phi\ddot{\theta} - \sin\theta\sin\phi\dot{\psi}\dot{\theta} + \cos\theta\cos\phi\dot{\psi}\dot{\phi} + \cos\theta\sin\phi\ddot{\psi} \tag{84}$$

$$\dot{r} = -\sin\theta\cos\phi\dot{\psi}\dot{\theta} - \cos\theta\sin\phi\dot{\psi}\dot{\phi} + \cos\theta\cos\phi\ddot{\psi} - \cos\phi\dot{\theta}\dot{\phi} - \sin\phi\ddot{\theta} \tag{85}$$

## InvSim flight path angles

The obtained explicit expressions in Eqs. (83–85) for the three body-referenced angular accelerations $(\dot{p}, \dot{q}, \dot{r})$ in the previous subsection contain the Euler angular velocities $(\dot{\phi}, \dot{\theta}, \dot{\psi})$ and the Euler angular accelerations $(\ddot{\phi}, \ddot{\theta}, \ddot{\psi})$. While $(\ddot{\phi})$ is considered available (either through numerically differentiating twice the input series values of $\phi$ or through evaluating a symbolic expression for $(\ddot{\phi})$ if $(\phi)$ is provided as a double-differentiable function of time), additional analytical expressions are needed for $(\ddot{\theta})$ and $(\ddot{\psi})$. These can be obtained by symbolically differentiating the two algebraic equations that relate the two flight path angles $(\psi_w, \theta_w)$ to the three Euler angles $(\phi, \theta, \psi)$ as well the angle of attack $(\alpha)$ and the sideslip angle $(\beta)$. These two algebraic equations to be differentiated are repeated below.

$$\cos\theta_w\sin(\psi_w - \psi) = \cos\phi\sin\beta - \sin\phi\sin\alpha\cos\beta \tag{33}$$

$$\sin\theta_w = \sin\theta\cos\alpha\cos\beta - \cos\theta\sin\phi\sin\beta - \cos\theta\cos\phi\sin\alpha\cos\beta \tag{34}$$

After performing symbolic differentiation twice for the above two algebraic equations, the resultant equations can be manipulated to obtain the sought explicit expressions for $(\ddot{\theta})$ and $(\ddot{\psi})$, as well as for the explicit expressions for the first derivatives $(\dot{\theta})$ and $(\dot{\psi})$. These $(\ddot{\theta})$ and $(\ddot{\psi})$ expressions are used in initializing the time loop for the numerical integration through the maneuver's discrete-time stations by estimating $(\dot{\theta}_{ini})$ and $(\dot{\psi}_{ini})$ at the start of the maneuver's simulation.

The symbolic explicit expressions for $(\dot{\phi}, \dot{\psi}, \ddot{\theta}, \ddot{\psi})$ are too complicated to be conveniently shown here. So, we alternatively refer to these four symbolic explicit expressions using the placeholder symbols $(\dot{\theta}_{\mathcal{F}}, \dot{\psi}_{\mathcal{F}}, \ddot{\theta}_{\mathcal{F}}$, and $\ddot{\psi}_{\mathcal{F}})$; respectively.

## InvSim linear-momentum equations and equilibrium

The first equation of the wind-referenced linear-momentum equations, which is Eq. (8), can be reformulated to be an explicit expression for the thrust force $(T)$, as follows:

$$T = \frac{1}{\cos\alpha\cos\beta}\left[m\dot{V} - \overline{q}S\left(C_x\,\cos\alpha\cos\beta + C_y\,\sin\beta + C_z\,\sin\alpha\cos\beta\right)\right.$$
$$\left. - mg\left(\cos\theta\sin\phi\sin\beta - \sin\theta\cos\alpha\cos\beta + \cos\theta\cos\phi\sin\alpha\cos\beta\right)\right] \tag{86}$$

This algebraic explicit expression for the thrust $(T)$ is used at the InvSim initialization stage for computing an initial thrust value $(T_{ini})$. Although the previous expression for $(T)$ has a singularity at $(\alpha = \pm 90°$ or $\pm \pi/2$ rad), these conditions are not realistic for a fixed-wing airplane, as they imply that the airplane is flying relatively perpendicular to its wing's surface. Also, although the previous expression for $(T)$ has a singularity at $(\beta = \pm 90°$ or $\pm \pi/2$ rad), these conditions are not realistic for a fixed-wing airplane as they imply that the airplane is drifting sideways by flying laterally, perpendicular to its longitudinal axis, similar to a cylinder interacting with a fluid flow and is oscillating laterally perpendicular to its longitudinal axis due to its interaction with the surrounding flow[347–352].

The explicit symbolic expressions $(\ddot{\theta}_{\mathcal{F}}$ and $\ddot{\psi}_{\mathcal{F}})$ discussed in the previous subsection involve the second time derivative of the angle of attack $(\ddot{\alpha})$ and the second time derivative of the sideslip angle $(\ddot{\beta})$, which appear in the right-hand side of $(\ddot{\theta}_{\mathcal{F}})$ and $(\ddot{\psi}_{\mathcal{F}})$. Therefore, two additional explicit symbolic expressions (to be denoted by the placeholder symbols $\ddot{\alpha}_{\mathcal{F}}$ and $\ddot{\beta}_{\mathcal{F}})$ are needed, which describe mathematically the second derivatives $(\ddot{\alpha}$ and $\ddot{\beta})$, respectively. Like $(\ddot{\theta}_{\mathcal{F}}$ and $\ddot{\psi}_{\mathcal{F}})$, the explicit expressions $(\ddot{\alpha}_{\mathcal{F}}$ and $\ddot{\beta}_{\mathcal{F}})$ are very elaborate and are not shown here, but we discuss next how these expressions can be obtained.

We provide below a reformulated version of the third equation of the wind-referenced linear momentum equations, which is Eq. (10) that represents the component along the third wind axis $(z_w)$, as an explicit expression for the first derivative of the angle of attack $(\dot{\alpha})$.

$$\dot{\alpha} = \frac{1}{mV\cos\beta}\left[\overline{q}S(C_z\,\cos\alpha - C_x\sin\alpha) + mg\,(\sin\theta\sin\alpha + \cos\,\theta\cos\,\phi\cos\alpha)\right.$$
$$\left. - T\sin\alpha + mV\,(q\cos\beta - r\sin\alpha\sin\beta - p\cos\alpha\sin\beta)\right] \tag{87}$$

Differentiating the above expression once with respect to time yields the explicit expression $(\ddot{\alpha}_{\mathcal{F}})$.

A reformulated version of the second equation of the wind-referenced linear momentum equations, which is Eq. (9), representing the linear momentum equation along the lateral wind axis $(y_w)$, is provided below as an explicit expression for the first derivative of the sideslip angle $(\dot{\beta})$.

$$\dot{\beta} = \frac{1}{mV} \left[ \bar{q}S \left( C_y \, \cos\beta - C_x \, \cos\alpha\sin\beta - C_z \, \sin\alpha\sin\beta \right) \right.$$
$$+ mg \left( \cos\theta\sin\phi\cos\beta + \sin\theta\cos\alpha\sin\beta - \cos\theta\cos\phi\sin\alpha\sin\beta \right)$$
$$\left. + T\cos\alpha\sin\beta + mV \left( -r\cos\alpha + p\sin\alpha \right) \right] \tag{88}$$

Differentiating the above expression once with respect to time yields the explicit expression $(\ddot{\beta}_\mathcal{F})$.

Although either of the two previous expressions for $(\dot{\alpha}$ and $\dot{\beta})$ has a singularity at $(V = 0)$, this condition is not realistic for a fixed-wing airplane, as it implies that the airplane is stagnant relative to the ambient air (a hovering condition).

In the explicit expressions $(\ddot{\alpha}_\mathcal{F}$ and $\ddot{\beta}_\mathcal{F})$, the first derivative of the thrust $(\dot{T})$ appears on the right-hand side. In order to construct a symbolic explicit expression for $(\dot{T})$; Eq. (86), which is an explicit expression for $(T)$ is differentiated once with respect to time. The resultant expression is very elaborate and thus is not shown, but is denoted by the placeholder symbol $(\dot{T}_\mathcal{F})$.

The discussions given before in "Linear-momentum equations and equilibrium" section and "Aerodynamic and stability coefficients" section regarding the interpretation of the angle of attack $(\alpha)$ appearing in the general flight mechanics modeling, and the relation of this angle to the conventional angle of attack $(\widehat{\alpha})$ and its equilibrium value $(\widehat{\alpha}_{equb})$ remain the same here for the InvSim mode of flight mechanics modeling.

### InvSim inertial velocity

Going back to the needed explicit expressions $(\ddot{\theta}_\mathcal{F}$ and $\ddot{\psi}_\mathcal{F})$ discussed earlier in "InvSim flight path angles" section; the explicit expression $(\ddot{\psi}_\mathcal{F})$ has the second time derivative of the azimuth flight path angle $(\ddot{\psi}_w)$ in its right-hand side, while the right-hand side of the explicit expression $(\ddot{\theta}_\mathcal{F})$ requires the second time derivative of both the azimuth flight path angle $(\ddot{\psi}_w)$ and the elevation flight path angle $(\ddot{\theta}_w)$.

The original expression for the inertial Cartesian velocity components $(\dot{x}_g, \dot{y}_g, \dot{z}_g)$ in "Inertial velocity" section can be mathematically altered to give two explicit expressions for these flight path angles $(\psi_w$ and $\theta_w)$. We recall below the original equations for the ground-referenced velocity components.

$$\dot{x}_g = V\cos\theta_w\cos\psi_w \tag{30}$$

$$\dot{y}_g = V\cos\theta_w\sin\psi_w \tag{31}$$

$$\dot{z}_g = -V\sin\theta_w \tag{32}$$

From these equations, the velocity of the airplane's center of gravity can be expressed in terms of the spherical flight-path coordinates $(V, \theta_w, \psi_w)$ as

$$V = \sqrt{\dot{x}_g^2 + \dot{y}_g^2 + \dot{z}_g^2} \tag{89}$$

$$\psi_w = \tan^{-1}\left(\frac{\dot{y}_g}{\dot{x}_g}\right) \tag{90}$$

$$\theta_w = \tan^{-1}\left(\frac{-\dot{z}_g}{\sqrt{\dot{x}_g^2 + \dot{y}_g^2}}\right) \tag{91}$$

The quantity $\left(\sqrt{\dot{x}_g^2 + \dot{y}_g^2}\right)$ is the projected velocity component in the horizon plane $(x_g - y_g)$, and it is equivalent to $(V\cos\theta_w)$ according to Eqs. (30 and 31). It can be shown that this quantity is also equivalent to $(\dot{x}_g\cos\psi_w + \dot{y}_g\sin\psi_w)$, given that $(\cos\psi_w = \dot{x}_g/\sqrt{\dot{x}_g^2 + \dot{y}_g^2})$ and $(\sin\psi_w = \dot{y}_g/\sqrt{\dot{x}_g^2 + \dot{y}_g^2})$.

It should also be noted that the derivatives of the inertial Cartesian coordinates $(\dot{x}_g, \dot{y}_g, \dot{z}_g)$ are computed either numerically using approximated finite difference expressions that are applied to the input discrete series of the inertial Cartesian coordinates $(x_g, y_g, z_g)$, or through evaluating symbolic expressions if these coordinates are described symbolically as differentiable functions. In either case, these derivatives $(\dot{x}_g, \dot{y}_g, \dot{z}_g)$ are assumed to eventually become available as a series of discrete values covering the entire maneuver duration.

The numerical values of the second derivatives of the inertial Cartesian coordinates $(\ddot{x}_g, \ddot{y}_g, \ddot{z}_g)$ are needed to evaluate the symbolic expressions for the first derivatives of the two flight angles $(\dot{\psi}_w$ and $\dot{\theta}_w)$, while the numerical values of the third derivatives of these inertial Cartesian coordinates $(\dddot{x}_g, \dddot{y}_g, \dddot{z}_g)$ are needed to evaluate the symbolic expressions for the second derivatives of the two flight angles $(\ddot{\psi}_w$ and $\ddot{\theta}_w)$. Again, the values of $(\ddot{x}_g, \ddot{y}_g, \ddot{z}_g)$ and $(\dddot{x}_g, \dddot{y}_g, \dddot{z}_g)$ can be obtained through numerical differentiation or symbolic differentiation (followed by numerical substitution at each of the time stations along the maneuver trajectory), depending on how the input ground-referenced coordinates $(x_g, y_g, z_g)$ for the maneuver trajectory are described.

Similarly, expressions for the first and second derivative of the velocity magnitude $(\dot{V}$ and $\ddot{V})$ can be developed in terms of the inertial accelerations $(\ddot{x}_g, \ddot{y}_g, \ddot{z}_g)$ and inertial jerks $(\dddot{x}_g, \dddot{y}_g, \dddot{z}_g)$. Alternatively, one may compute the discrete values of $(V, \theta_w, \psi_w)$ from the numerical values of $(\dot{x}_g, \dot{y}_g, \dot{z}_g)$ by using the above Eqs. (89–91), and then applying finite difference formulas to compute the values of the first time derivatives $(\dot{V}, \dot{\theta}_w, \dot{\psi}_w)$ and the second time derivatives $(\ddot{V}, \ddot{\theta}_w, \ddot{\psi}_w)$.

We list below standard formulas of the finite difference method (FDM) to approximate a first derivative $(\dot{f})$, a second derivative $(\ddot{f})$, and a third derivative $(\dddot{f})$, of a generic time-dependent function $f(t)$, if its values at uniformly-spaced time stations having a fixed time step ($\Delta t$) are known[353–357]. These formulas are second-order-accurate, which means that the discretization error decays at a rate that is related to the decay rate of the time step squared. For each derivative order (one, two, or three); we provide three types of finite difference formulas, namely (1) forward difference (FD), (2) central difference (CD), and (3) backward difference (BD). At the initial time station (assigned a time station index $n = 1$), forward difference should be used, since no past "backward" values are available. At the final time station (assigned a time station index $n = n_{max}$), backward difference should be used, since no future "forward" values are available. At other intermediate time stations (having time station index values of $2 \leq n \leq n_{max} - 1$ for $\dot{f}$ and $\ddot{f}$; but $3 \leq n \leq n_{max} - 2$ for $\dddot{f}$), central difference is preferred because it involves less computation and it is symmetric (considering both past and future data). The finite difference formulas are listed in Table 8. In this table, the subscript indices (such as "$n - 1$", "$n$", and "$n + 1$") refer to the time position relative to the generic time station ($n$).

With this, the numerical values of the following 12 quantities should be known at each time station before the main InvSim flight mechanics simulation starts:

1. $V, \dot{V}, \ddot{V}$
2. $\theta_w, \dot{\theta}_w, \ddot{\theta}_w$
3. $\psi_w, \dot{\psi}_w, \ddot{\psi}_w$
4. $\phi, \dot{\phi}, \ddot{\phi}$

We provide below symbolic expressions for the first derivatives ($\dot{V}, \dot{\theta}_w, \dot{\psi}_w$), while the symbolic expressions for the second derivatives ($\ddot{V}, \ddot{\theta}_w, \ddot{\psi}_w$) are much more detailed and thus are not provided here, but we denote them by the placeholder symbols ($\ddot{V}_\mathcal{F}, \ddot{\theta}_{w,\mathcal{F}}, \ddot{\psi}_{w,\mathcal{F}}$), respectively.

$$\dot{V} = \frac{\dot{x}_g \ddot{x}_g + \dot{y}_g \ddot{y}_g + \dot{z}_g \ddot{z}_g}{\sqrt{\dot{x}_g^2 + \dot{y}_g^2 + \dot{z}_g^2}} = \frac{\dot{x}_g \ddot{x}_g + \dot{y}_g \ddot{y}_g + \dot{z}_g \ddot{z}_g}{V} \tag{92}$$

$$\dot{\psi}_w = \frac{\ddot{y}_g \dot{x}_g - \ddot{x}_g \dot{y}_g}{\dot{x}_g^2 + \dot{y}_g^2} = \frac{\ddot{y}_g \cos(\psi_w) - \ddot{x}_g \sin(\psi_w)}{\dot{x}_g \cos(\psi_w) + \dot{y}_g \sin(\psi_w)} \tag{93}$$

$$\dot{\theta}_w = \frac{\frac{-\ddot{z}_g}{\sqrt{\dot{x}_g^2 + \dot{y}_g^2}} + \frac{\dot{z}_g (\dot{x}_g \ddot{x}_g + \dot{y}_g \ddot{y}_g)}{(\dot{x}_g^2 + \dot{y}_g^2)^{1.5}}}{1 + \frac{\dot{z}_g^2}{\dot{x}_g^2 + \dot{y}_g^2}} = -\frac{\ddot{z}_g + \dot{V} \sin(\theta_w)}{V \cos(\theta_w)} \tag{94}$$

Either of the two previous expressions for ($\dot{\psi}_w$) and ($\dot{\theta}_w$) has a singularity at ($\dot{x}_g = \dot{y}_g = 0$), which corresponds to ($\theta_w = \pm 90°$ or $\pm \pi/2$ rad). In a real flight setting, this means that the airplane is flying straight up or straight down with respect to the fixed ground (thus, the airplane is flying perpendicular to the horizon). This is a restriction in the present InvSim numerical algorithm, which fails in this exceptional case. Therefore, such two particular flight movements (vertical ascent, nose-up; and vertical descent, nose-down) should be excluded, although some military airplanes are capable of performing such unconventional flight situations, with their thrust exceeding their weight[358–360]. For civil airplanes and military airplanes not performing this extreme maneuver, this restriction is not a concern.

## InvSim three aerodynamic forces

In the proposed InvSim here for the inverse simulation of flight mechanics problems, the nondimensional aerodynamic coefficients along the body axes ($C_x, C_y, C_z$) are used in lieu of the dimensional aerodynamic forces along the body axes ($X, Y, Z$). Therefore, the explicit expressions for the aerodynamic forces presented in the original flight mechanics formulation, Eqs. (35–37) in "Three aerodynamic forces" section, are not needed for performing the InvSim computations. However, these expressions can still be used for computing and reporting these aerodynamic forces as supplementary post-processing quantities.

| Derivative | Difference type | Expression |
|---|---|---|
| $\dot{f}$ | Forward | $\dot{f}_n \cong \frac{1}{2\Delta t}(-3f_n + 4f_{n+1} - f_{n+2})$ |
| $\dot{f}$ | Central | $\dot{f}_n \cong \frac{1}{2\Delta t}(f_{n+1} - f_{n-1})$ |
| $\dot{f}$ | Backward | $\dot{f}_n \cong \frac{1}{2\Delta t}(3f_n - 4f_{n-1} + f_{n-2})$ |
| $\ddot{f}$ | Forward | $\ddot{f}_n \cong \frac{1}{\Delta t^2}(2f_n - 5f_{n+1} + 4f_{n+2} - f_{n+3})$ |
| $\ddot{f}$ | Central | $\ddot{f}_n \cong \frac{1}{\Delta t^2}(f_{n+1} - 2f_n + f_{n-1})$ |
| $\ddot{f}$ | Backward | $\ddot{f}_n \cong \frac{1}{\Delta t^2}(2f_n - 5f_{n-1} + 4f_{n-2} - f_{n-3})$ |
| $\dddot{f}$ | Forward | $\dddot{f}_n \cong \frac{1}{2\Delta t^3}(-5f_n + 18f_{n+1} - 24f_{n+2} + 14f_{n+3} - 3f_{n+4})$ |
| $\dddot{f}$ | Central | $\dddot{f}_n \cong \frac{1}{2\Delta t^3}(f_{n+2} - 2f_{n-1} + 2f_{n-1} - f_{n-2})$ |
| $\dddot{f}$ | Backward | $\dddot{f}_n \cong \frac{1}{2\Delta t^3}(5f_n - 18f_{n-1} + 24f_{n-2} - 14f_{n-3} + 3f_{n-4})$ |

**Table 8.** General finite difference formulas for numerical differentiation.

This optionality is similar to the one discussed earlier in subsection 3.9 for auxiliary quantities such as the speed of sound ($a$).

## InvSim air density and speed of sound

The variation of the air density ($\rho$), air temperature ($\Theta$), and speed of sound ($a$) with the altitude ($h$) as described by the International Standard Atmosphere (ISA) model and discussed earlier for the general flight mechanics formulation are independent of the mode of solving the flight mechanics problem (either forward mode or inverse mode). Rather, the dependence of these three air properties on the altitude is described by standalone relationships that serve to supply the proper air conditions at the altitude of each time station.

Therefore, this part of the InvSim algorithm is not repeated here, as it is the same as the one discussed earlier for the general flight mechanics formulation in "Air density and speed of sound" section.

We clarify again that only the air density is needed in the InvSim algorithm computations, while the air temperature only facilitates computing the local speed of sound in air, which in turn permits computing the Mach number as a supplementary flight feature.

## InvSim numerical algorithm

In this section, we describe the main stages of the numerical integration algorithm for solving the inverse simulation (InvSim) flight mechanics problem for an airplane, based on the miscellaneous equations and theoretical overview given before.

In summary, the algorithm has three main stages, namely (1) pre-processing of inputs, (2) initialization, and (3) time loop Runge–Kutta method.

### Pre-processing of Inputs

This first stage of the InvSim numerical algorithm can further be divided into the following 17 steps:

1. The initial time for the trajectory is set to ($t = 0$). If the entire trajectory duration is ($t_{max}$), then the total number of time stations is

$$n_{max} = \frac{t_{max}}{\Delta t} + 1 \qquad (95)$$

where ($\Delta t$) is the uniform time step; and the number of time steps (the transitions between successive time stations) is ($n_{max} - 1$) or ($t_{max}/\Delta t$).

2. The 30 constants needed for defining the airplane's geometry and its aerodynamic/stability behavior, as well as the initial altitude (as listed in "Summary of constants" section), are received by the user.
3. The four main InvSim inputs are also received by the user, either as analytical (symbolic) expressions or as equally-spaced discrete values (a quantitative vector) with a constant time step ($\Delta t$). These four main inputs to the InvSim algorithm are

- Ground-referenced inertial coordinates of the maneuver trajectory: $x_g, y_g, z_g$
- roll angle: $\phi$

4. Using the initial altitude ($h_{ini}$) and the values of ($z_g$) into Eq. (1), all numerical values of the flight altitude ($h$) are obtained.
5. Using the obtained values of the altitude ($h$) into either Eq. (62) for the troposphere layer or Eq. (66) for the tropopause layer, all numerical values of the air density ($\rho$) are obtained.

Optionally, all numerical values of the speed of sound can also be computed as we explained in "Air density and speed of sound" section.

6. Using either analytical (symbolic) differentiation (if the roll angle $\phi$ is provided in a functional form) or second-order finite difference formulas (if the roll angle $\phi$ is provided as a vector of discrete values), all numerical values of the first derivative of the roll angle ($\dot{\phi}$) and the second derivative of the roll angle ($\ddot{\phi}$) are obtained.
7. Using either analytical (symbolic) differentiation (if the inertial coordinates $x_g, y_g, z_g$ are provided in a functional form) or second-order finite difference formulas (if the inertial coordinates $x_g, y_g, z_g$ are provided as vectors of discrete values); all numerical values of their first derivatives ($\dot{x}_g, \dot{y}_g, \dot{z}_g$), second derivatives ($\ddot{x}_g, \ddot{y}_g, \ddot{z}_g$), and third derivatives ($\dddot{x}_g, \dddot{y}_g, \dddot{z}_g$) are obtained.
8. Using Eq. (89), all numerical values of the velocity magnitude ($V$) are obtained.
9. Using the obtained values of the air density ($\rho$) and the obtained values of the airplane speed ($V$) in Eq. (11), all numerical values of the dynamic pressure ($\overline{q}$) are obtained.
10. Using Eq. (90), all numerical values of the azimuth flight path angle ($\psi_w$) are obtained.
11. Using Eq. (91), all numerical values of the elevation flight path angle ($\theta_w$) are obtained.
12. Using second-order finite difference formulas discussed in "InvSim inertial velocity" section, all numerical values of the first derivative ($\dot{V}$) are obtained.

Alternatively, Eq. (92) can be used.

13. Using the Eq. (93), all numerical values of the first derivative ($\dot{\psi}_w$) are obtained.
14. Using the Eq. (94), all numerical values of the first derivative ($\dot{\theta}_w$) are obtained.
15. Using second-order finite difference formulas discussed in "InvSim inertial velocity" section, all numerical values of the second derivative ($\ddot{V}$) are obtained.

Alternatively, the explicit expression ($\ddot{V}_{\mathcal{F}}$) discussed in "InvSim inertial velocity" section, can be used.

16. Using the explicit expression ($\ddot{\psi}_{w,\mathcal{F}}$) discussed in "InvSim inertial velocity" section, all numerical values of the second derivative ($\ddot{\psi}_w$) are obtained.
17. Using the explicit expression ($\ddot{\theta}_{w,\mathcal{F}}$) discussed in "InvSim inertial velocity" section, all numerical values of the second derivative ($\ddot{\theta}_w$) are obtained.

At the end of this first stage of the InvSim numerical algorithm, the following 27 quantitative vectors (one-dimensional arrays, or series of numerical values) become available, covering all the ($n_{max}$) time stations of the flight maneuver:

1. $x_g, y_g, z_g, \phi$ ( the four InvSim inputs)
2. $\dot{x}_g, \dot{y}_g, \dot{z}_g, \dot{\phi}$
3. $\ddot{x}_g, \ddot{y}_g, \ddot{z}_g, \ddot{\phi}$
4. $\dddot{x}_g, \dddot{y}_g, \dddot{z}_g$
5. $V, \dot{V}, \ddot{V}$
6. $h, \rho, \overline{q}$
7. $\theta_w, \dot{\theta}_w, \ddot{\theta}_w$
8. $\psi_w, \dot{\psi}_w, \ddot{\psi}_w$

Optionally, four more pre-processing quantitative vectors can be computed entirely from the user's provided data, before solving for other flight variables; although these four quantities are not core elements within the InvSim algorithm, thus the simulation can be completed without knowing these quantities. These four optional pre-processing vectors are

9. $\Theta, a, M, \widetilde{\alpha}_{equb}$

## Initialization
The initial conditions correspond to the first time station, with the index ($n = 1$). The following initial conditions are implemented:

1. The initial angle of attack ($\alpha_{ini}$) and the initial sideslip angle ($\beta_{ini}$) are set to zero, in alignment with the equilibrium condition.
2. The above-mentioned initial zero values for ($\alpha_{ini}$ and $\beta_{ini}$) dictate that the initial values of the two unspecified Euler angles ($\psi_{ini}$ and $\theta_{ini}$) take the same initial values of the known spherical flight path angles ($\psi_w$ and $\theta_w$), respectively. Therefore, $\psi_{ini} = \psi_w(n = 1)$ and $\theta_{ini} = \theta_w(n = 1)$. These initial conditions are implied by Eqs. (33) and (34), respectively, as discussed in "Flight path angles" section.
3. The initial thrust force ($T_{ini}$) is computed using Eq. (86).
4. The initial angle of attack rate ($\dot{\alpha}_{ini}$) and the initial sideslip angle rate ($\dot{\beta}_{ini}$) are set to zero.
5. The explicit expressions ($\dot{\psi}_{\mathcal{F}}$ and $\dot{\theta}_{\mathcal{F}}$) discussed in "InvSim flight path angles" section imply that the initial Euler yaw rate ($\dot{\psi}_{ini}$) and the initial Euler pitch rate ($\dot{\theta}_{ini}$), respectively, have zero values.
6. Equations (2, 3, 4) imply that the initial body-referenced angular velocities ($p_{ini}, q_{ini}, r_{ini}$) respectively, have zero values.
7. Equations (83, 84, 85) imply that the initial body-referenced angular acceleration ($\dot{p}_{ini}, \dot{q}_{ini}, \dot{r}_{ini}$), respectively, have zero values.
8. The initial control surface deflection angles ($\delta_{m,ini}, \delta_{l,ini}, \delta_{n,ini}$) are computed using Eqs. (73, 74, 75), respectively.

## Time loop Runge–Kutta method
This third stage of the InvSim numerical algorithm is the main stage, where the classical explicit fourth-order Runge–Kutta (RK4) four-step integration method is used to numerically compute the evolution of eight flight variables through a system of nonlinear coupled ordinary differential equations (ODE)[361–365]. These eight variables are:

- $T$
- $\alpha, \beta$
- $\psi, \theta$
- $p, q, r$

In this subsection, we describe how these eight variables are advanced in discrete time through the proposed InvSim numerical algorithm from an arbitrary time station (discrete time index $n$) with known values for these eight variables to the next time station (discrete time index $n + 1$) with unknown values for these eight variables.

In the beginning, it might be useful to make a brief description of the classical fourth-order Runge–Kutta method for a generic nonlinear scalar ordinary differential equation (ODE) of the form

$$\dot{y}(t, y) = f(t, y) \tag{96}$$

with a known starting condition denoted as $y_{old}(t_{old})$.

The procedure of numerically integrating this exemplar ODE in order to compute the value $y_{new}(t_{old} + \Delta t)$ is as follows:

$$
\begin{aligned}
&\dot{y}_1^* = f\left(t_{old}, y_{old}\right); t_1^* = t_{old} + \frac{\Delta t}{2}; y_1^* = y_{old} + \frac{\Delta t}{2}\dot{y}_1^* \\
&\dot{y}_2^* = f\left(t_1^*, y_1^*\right); t_2^* = t_1^* = t_{old} + \frac{\Delta t}{2}; y_2^* = y_{old} + \frac{\Delta t}{2}\dot{y}_2^* \\
&\dot{y}_3^* = f\left(t_2^*, y_2^*\right); t_3^* = t_{old} + \Delta t; y_3^* = y_{old} + \Delta t\dot{y}_3^* \\
&\dot{y}_4^* = f\left(t_3^*, y_3^*\right) \\
&y_{new}\left(t_{new} = t_{old} + \Delta t\right) = y_{old} + \frac{\Delta t}{6}\left(\dot{y}_1^* + 2\dot{y}_2^* + 2\dot{y}_3^* + \dot{y}_4^*\right)
\end{aligned}
\tag{97}
$$

where the superscript asterisk in ($\dot{y}^*$, $t^*$, and $y^*$) refers to a temporary (intermediate) value.

Considering the proposed InvSim algorithm presented here, the above procedure is to be applied four times per time step as follows:

1. Either of the four intermediate derivatives for the thrust ($\dot{T}_1^*, \dot{T}_2^*, \dot{T}_3^*, \dot{T}_4^*$) is computed using the explicit expression ($\dot{T}_{\mathcal{F}}$) discussed in "InvSim linear-momentum equations and equilibrium" section.

2. Using the explicit expressions ($\ddot{\alpha}_{\mathcal{F}}$ and $\ddot{\beta}_{\mathcal{F}}$) along with Eqs. (87 and 88) discussed in "InvSim linear-momentum equations and equilibrium" section, the intermediate derivatives for the angle of attack ($\dot{\alpha}_1^*, \dot{\alpha}_2^*, \dot{\alpha}_3^*, \dot{\alpha}_4^*$) and the sideslip angle ($\dot{\beta}_1^*, \dot{\beta}_2^*, \dot{\beta}_3^*, \dot{\beta}_4^*$) are computed, respectively as

$$\dot{\alpha}_1^* = \dot{\alpha}_{old}; \dot{\beta}_1^* = \dot{\beta}_{old}$$

$$\dot{\alpha}_2^* = \dot{\alpha}_{old} + \frac{\Delta t}{2}\ddot{\alpha}_1^*; \dot{\beta}_2^* = \dot{\beta}_{old} + \frac{\Delta t}{2}\ddot{\beta}_1^*$$

$$\dot{\alpha}_3^* = \dot{\alpha}_{old} + \frac{\Delta t}{2}\ddot{\alpha}_2^*; \dot{\beta}_3^* = \dot{\beta}_{old} + \frac{\Delta t}{2}\ddot{\beta}_2^*$$

$$\dot{\alpha}_4^* = \dot{\alpha}_{old} + \Delta t\ddot{\alpha}_3^*; \dot{\beta}_4^* = \dot{\beta}_{old} + \Delta t\ddot{\beta}_3^* \tag{98}$$

3. Using the explicit expressions ($\ddot{\psi}_{\mathcal{F}}$ and $\ddot{\theta}_{\mathcal{F}}$) along with the explicit expressions ($\dot{\psi}_{\mathcal{F}}$ and $\dot{\theta}_{\mathcal{F}}$) discussed in "InvSim flight path angles" section, the intermediate derivatives ($\dot{\psi}_1^*, \dot{\psi}_2^*, \dot{\psi}_3^*, \dot{\psi}_4^*$) and ($\dot{\theta}_1^*, \dot{\theta}_2^*, \dot{\theta}_3^*, \dot{\theta}_4^*$) are computed as

$$\dot{\psi}_1^* = \dot{\psi}_{old}; \dot{\theta}_1^* = \dot{\theta}_{old}$$

$$\dot{\psi}_2^* = \dot{\psi}_{old} + \frac{\Delta t}{2}\ddot{\psi}_1^*; \dot{\theta}_2^* = \dot{\theta}_{old} + \frac{\Delta t}{2}\ddot{\theta}_1^*$$

$$\dot{\psi}_3^* = \dot{\psi}_{old} + \frac{\Delta t}{2}\ddot{\psi}_2^*; \dot{\theta}_3^* = \dot{\theta}_{old} + \frac{\Delta t}{2}\ddot{\theta}_2^*$$

$$\dot{\psi}_4^* = \dot{\psi}_{old} + \Delta t\ddot{\psi}_3^*; \dot{\theta}_4^* = \dot{\theta}_{old} + \Delta t\ddot{\theta}_3^* \tag{99}$$

4. Either of the four intermediate derivatives of the body-referenced roll rate ($\dot{p}_1^*, \dot{p}_2^*, \dot{p}_3^*, \dot{p}_4^*$) is computed using Eq. (83). Similarly, either of the four intermediate derivatives of the body-referenced pitch rate ($\dot{q}_1^*, \dot{q}_2^*, \dot{q}_3^*, \dot{q}_4^*$) is computed using Eq. (84), and either of the four intermediate derivatives of the body-referenced yaw rate ($\dot{r}_1^*, \dot{r}_2^*, \dot{r}_3^*, \dot{r}_4^*$) is computed using Eq. (85).

5. The computed 32 temporary derivatives (eight temporary derivatives for eight variables are computed in each of the four steps of the RK4 procedure) are used to update the eight flight variables of the RK4 procedure at time station ($n + 1$), as follows:

$$T_{new}\left(\text{station}:n+1\right) = T_{old}\left(\text{station}:n\right) + \frac{\Delta t}{6}\left(\dot{T}_1^* + 2\dot{T}_2^* + 2\dot{T}_3^* + \dot{T}_4^*\right)$$

$$\alpha_{new}\left(\text{station}:n+1\right) = \alpha_{old}\left(\text{station}:n\right) + \frac{\Delta t}{6}\left(\dot{\alpha}_1^* + 2\dot{\alpha}_2^* + 2\dot{\alpha}_3^* + \dot{\alpha}_4^*\right)$$

$$\beta_{new}\left(\text{station}:n+1\right) = \beta_{old}\left(\text{station}:n\right) + \frac{\Delta t}{6}\left(\dot{\beta}_1^* + 2\dot{\beta}_2^* + 2\dot{\beta}_3^* + \dot{\beta}_4^*\right)$$

$$\psi_{new}\left(\text{station}:n+1\right) = \psi_{old}\left(\text{station}:n\right) + \frac{\Delta t}{6}\left(\dot{\psi}_1^* + 2\dot{\psi}_2^* + 2\dot{\psi}_3^* + \dot{\psi}_4^*\right)$$

$$\theta_{new}\left(\text{station}:n+1\right) = \theta_{old}\left(\text{station}:n\right) + \frac{\Delta t}{6}\left(\dot{\theta}_1^* + 2\dot{\theta}_2^* + 2\dot{\theta}_3^* + \dot{\theta}_4^*\right) \quad (100)$$

$$p_{new}\left(\text{station}:n+1\right) = p_{old}\left(\text{station}:n\right) + \frac{\Delta t}{6}\left(\dot{p}_1^* + 2\dot{p}_2^* + 2\dot{p}_3^* + \dot{p}_4^*\right)$$

$$q_{new}\left(\text{station}:n+1\right) = q_{old}\left(\text{station}:n\right) + \frac{\Delta t}{6}\left(\dot{q}_1^* + 2\dot{q}_2^* + 2\dot{q}_3^* + \dot{q}_4^*\right)$$

$$r_{new}\left(\text{station}:n+1\right) = r_{old}\left(\text{station}:n\right) + \frac{\Delta t}{6}\left(\dot{r}_1^* + 2\dot{r}_2^* + 2\dot{r}_3^* + \dot{r}_4^*\right)$$

With this update, one of the sought four output flight controls (namely the thrust, $T$) becomes known at the new time station $(n+1)$.

We need to obtain the remaining three sought output flight controls (the three moving surface deflection angles; $\delta_m, \delta_l, \delta_n$). This is done in the remaining part of the InvSim numerical algorithm as explained next.

6. The new values (at time station $n+1$) of the eight updated flight variables through the RK4 procedure are used to evaluate several derivative expressions, ending with the body-referenced angular acceleration $(\dot{p}, \dot{q}, \dot{r})$ at the new time station $(n+1)$ using Eqs. (83, 84, 85), respectively.

7. The obtained new body-referenced angular accelerations $(\dot{p}, \dot{q}, \dot{r})$ are used in the explicit algebraic expressions $(T_{1,\mathcal{F}}, T_{2,\mathcal{F}}, T_{3,\mathcal{F}})$ to compute the auxiliary moments $(T_1, T_2, T_3)$ at the new time station $(n+1)$.

8. The obtained new auxiliary moments $(T_1, T_2, T_3)$ are used in Eqs. (80, 81, 82), respectively, to find the corresponding total body-referenced moments $(L, M, N)$ at the new time station $(n+1)$.

9. The obtained new total body-referenced moments $(L, M, N)$ are used in Eqs. (76, 77, 78), respectively, to find the corresponding nondimensional moment coefficients $(C_l, C_m, C_n)$ at the new time station $(n+1)$.

10. Finally, the obtained new nondimensional moment coefficient $(C_m)$ is used in Eq. (73) to find the necessary deflection angle for the elevators $(\delta_m)$ at the new time station $(n+1)$. Similarly, the obtained new nondimensional moment coefficients $(C_l, C_n)$ are used in Eqs. (74, 75), respectively, to find the necessary deflection angle for the ailerons $(\delta_l)$ and the necessary deflection for the rudder $(\delta_n)$ at the new time station $(n+1)$.

If this is not the last time station $(n \neq n_{max})$, the above 10 computational actions are repeated to find the four necessary flight control variables $(T, \delta_m, \delta_l, \delta_n)$ at the subsequent time station. The latest new values obtained at $(n+1)$ become old "known" values when advancing from the "known" time station $(n+1)$ to the "unknown" $(n+2)$.

## InvSim example for Mirage III

Although the main contribution of this work is the detailed presentation of the mathematical formulation and the numerical algorithm for inverse simulation (InvSim) flight mechanics, where the required flight controls corresponding to a desired flight trajectory for a given fixed-wing airplane may be computed; it is a valuable addition to this study to provide an example case in which the proposed InvSim algorithm is applied. This demonstrative example is performed in this section for a set of airplane data representing the family of military fighters called "Mirage III".

### About Mirage III

Mirage III is a family of military all-weather aircraft capable of interception at supersonic speeds and capable of taking off using compact runways. Mirage III airplanes are produced by the French aerospace representation Dassault Aviation, which is a major partner of the French national defense[366–370]. The Mirage III family is characterized by a delta wing (shaped as a triangle), and a single powerful jet engine able to deliver a thrust force of about 80 kN (80,000 N).

Figure 9 is a photo taken for the model (V 01) of Mirage III, which provides a general overview of its geometric design[371].

Although there are different models of Mirage III, we here use the following set of geometric and performance parameters listed in Table 9, which are considered to be an archetype representation of Mirage III. Due to the delta-shaped wing, the characteristic longitudinal length $(c)$ and the characteristic lateral length $(b)$ are assigned equal values.

The above value of the wing area is gross, thus it includes part of the projected fuselage area located between the wing tips. The equality of the span and chord in the above table is an unusual feature of the tailless Mirage fighter airplanes, given their delta wing having a steep rearward sweep angle and a low aspect ratio near unity.

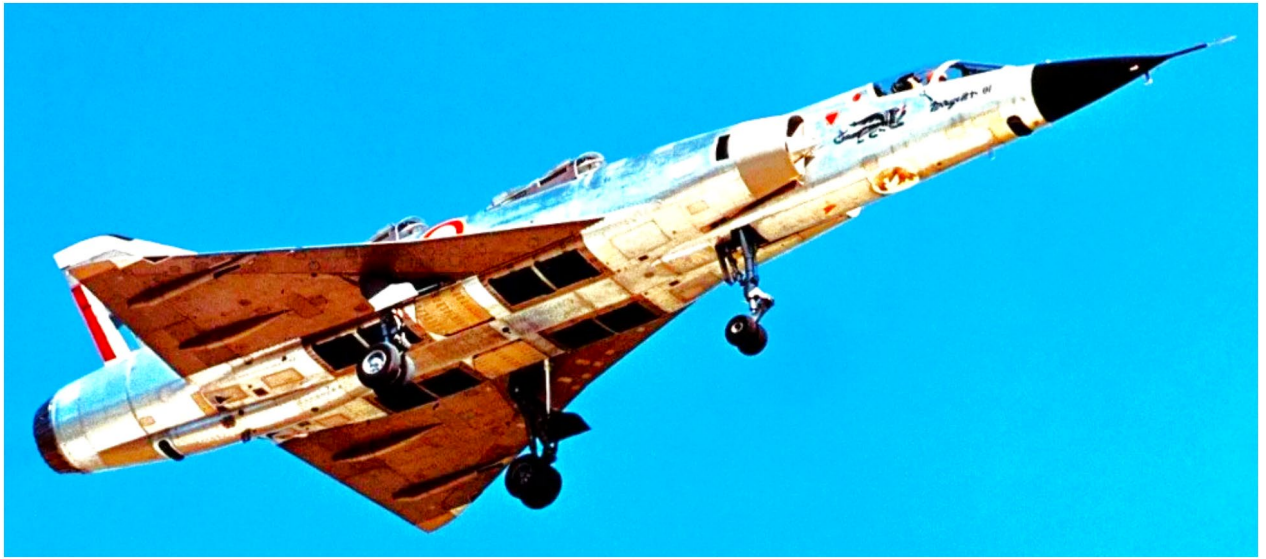

**Fig. 9**. Photo of Mirage III (model V 01) in flight. Permission for use was requested from the manufacturer Dassault Aviation.

| Serial number | Parameter | Value |
|---|---|---|
| 1 | Mass ($m$) | 7,400 kg |
| 2 | Wing planform area ($S$) | 36 m² |
| 3 | Mean aerodynamic chord ($c$) | 5.25 m |
| 4 | Wing span ($b$) | 5.25 m |
| 5 | Moment of inertia about $x_b$ ( $A$) | 90,000 kg.m² |
| 6 | Moment of inertia about $y_b$ ( $B$) | 54,000 kg.m² |
| 7 | Moment of inertia about $z_b$ ( $C$) | 60,000 kg.m² |
| 8 | Moment of inertia in $y_b - z_b$ ( $D$) | 0 kg.m² |
| 9 | Moment of inertia in $x_b - z_b$ ( $E$) | 1,800 kg.m² |
| 10 | Moment of inertia in $x_b - y_b$ ( $F$) | 0 kg.m² |
| 11 | Lift coefficient at zero conventional angle of attack ($C_{L0}$) | 0 |
| 12 | Slope of lift coefficient ($C_{L\alpha}$) | 2.204 1/rad |
| 13 | Drag coefficient at zero lift ($C_{D0}$) | 0.015 |
| 14 | Drag polar parameter ($K_{CD}$) | 0.4 |
| 15 | Side-force coefficient parameter ($C_{C\beta}$) | –0.6 1/rad |
| 16 | Longitudinal stability parameter ($C_{m0}$) | 0 |
| 17 | Longitudinal stability parameter ($C_{m\alpha}$) | –0.17 |
| 18 | Longitudinal stability parameter ($C_{mq}$) | –0.4 |
| 19 | Longitudinal stability parameter ($C_{m\delta m}$) | –0.45 1/rad |
| 20 | Lateral stability parameter ($C_{l\beta}$) | –0.05 1/rad |
| 21 | Lateral stability parameter ($C_{lp}$) | –0.25 |
| 22 | Lateral stability parameter ($C_{lr}$) | 0.06 |
| 23 | Lateral stability parameter ($C_{l\delta l}$) | –0.3 |
| 24 | Lateral stability parameter ($C_{l\delta n}$) | 0.018 1/rad |
| 25 | Directional stability parameter ($C_{n\beta}$) | 0.15 1/rad |
| 26 | Directional stability parameter ($C_{np}$) | 0.055 |
| 27 | Directional stability parameter ($C_{nr}$) | –0.7 |
| 28 | Directional stability parameter ($C_{n\delta l}$) | 0 |
| 29 | Directional stability parameter ($C_{n\delta n}$) | –0.085 1/rad |

**Table 9**. Adopted geometric parameters for Mirage III.

In addition, we select an initial altitude of $h_{ini} = 0$. This means that the trajectory is assumed to start from a point located at the mean sea level, and this is a reasonable assumption, and it helps in standardizing our proposed maneuver as a numerical testing case in flight mechanics.

## Proposed test maneuver

The flight maneuver we propose here to verify the applicability of the presented inverse simulation (InvSim) numerical algorithm is a simple one, being a horizontal north-wise straight flight at a constant altitude of 5,000 m and with a constant flight speed of 150 m/s while performing a continuous double-roll rotation with a non-linear angular profile constituting of a two-harmonic function. This simple maneuver has several advantages, making it a favored test not only in the current study but also in general for the flight mechanics modeling community globally, including those who want to develop the presented InvSim algorithm, and test its implementation.

With this proposed test maneuver, it is easier to inspect the computational procedure of various flight variables than with a complex maneuver. In our proposed maneuver, different vector quantities reduce practically to an array of a single value repeated over all time stations. In addition, various derivatives become zero throughout the entire maneuver. In addition, all the main InvSim input profiles can be specified through concise analytical (symbolic) functions, rather than as extensive numerical lists.

The four main inputs to the InvSim flight mechanics algorithm in our test maneuver are

$$
\begin{aligned}
x_g(t) &= 150t \\
y_g &= 0 \\
z_g &= -5{,}000 \\
\phi(t) &= \frac{\pi}{4}\left[8 + \cos\left(\frac{\pi}{10}t\right) - 9\cos\left(\frac{\pi}{30}t\right)\right]
\end{aligned}
\tag{101}
$$

where $(x_g, y_g, z_g)$ are in meters, and $(\phi)$ is in radians.

Taking the derivatives of the roll angle $(\phi)$ gives the following expressions for its Euler rate $(\dot{\phi})$, and Euler acceleration $(\ddot{\phi})$:

$$
\begin{aligned}
\dot{\phi}(t) &= \frac{\pi^2}{40}\left[-\sin\left(\frac{\pi}{10}t\right) + 3\sin\left(\frac{\pi}{30}t\right)\right] \\
\ddot{\phi}(t) &= \frac{\pi^3}{400}\left[-\cos\left(\frac{\pi}{10}t\right) + \cos\left(\frac{\pi}{30}t\right)\right]
\end{aligned}
\tag{102}
$$

Figures 10, 11, and 12 illustrate the temporal profile of the roll angle $(\phi)$, its first-time derivative $(\dot{\phi})$, and its second-time derivative $(\ddot{\phi})$; respectively, during the proposed double-roll maneuver. In these figures, the angles and their gradients are expressed in degrees, degrees per second, or degrees per square second (through scaling by the multiplicative value $180/\pi = 57.296°$/rad) for easier comprehension since the degree unit is broadly used. The roll angle changes from 0° to 720° over a duration of 30 s. The rate $(\dot{\phi})$ has a non-negative symmetric profile, peaking at the midpoint of 15 s to reach a maximum value of 56.549°/s; while $(\ddot{\phi})$ has an antisymmetric profile, with a positive peak of 6.838°/s² at 9.123 s and a negative peak of $-6.838°$/s² at 20.877 s.

Table 10 lists some of the computed characteristics for the test maneuver. These characteristics aid in validating any developed computational implementations by other researchers because these quantities can be computed at the initial time station ($n = 1$, $t = 0$) independently of the time loop calculations. In addition, these quantitative characteristics provide useful insights into the nature of this numerical test maneuver when viewed as a real flight mission. It should be noted that due to the maneuver's simplicity, these scalar quantities are actually vectors of constants (vectors or arrays representing frozen or static variables), thus all the time ($n_{max}$) stations have the same value for each characteristic quantity.

## Inverse simulation results

We implemented the proposed InvSim numerical algorithm as a MATLAB/Octave computer code, and we applied it here using GNU Octave version 6.1.0[372]. The algorithm does not require any specialized toolboxes, which is a big advantage of simplicity. The time step ($\Delta t$) we used is 0.001 s (thus, there is a total of $n_{max} = 30{,}001$ time stations). We tried different values for the time step, and we found that this selected value is proper. With much lower values (such as 0.01 s), the simulation can still be performed without instability, but minor spurious oscillations or non-smooth variations near sharp peaks may occur.

We start the results of the InvSim algorithm for the test maneuver with the four outputs, which are the flight controls. Figure 13 shows the computed thrust force, which is one of the four flight controls. The equilibrium value is 11,543 N (11.543 kN), which changes between four minima and three maxima until the same equilibrium value is restored at the end of the maneuver. The three encountered maxima values of the thrust are 11,332 N at 11.613 s; 11,535 N at 15.002 s; and 11,348 N at 18.390 s. All these peaks are less than (but close to) the equilibrium value. Knowing the global maximum thrust needed during a prospective maneuver is particularly important because it helps in deciding whether that target maneuver is achievable or not, based on the available maximum thrust for the airplane of concern. The four thrust minima are nearly 4,900 N. Knowing the global minimum thrust needed during a maneuver is also important, because negative values (reversed backward thrust) typically imply that the maneuver is not achievable.

Figure 14 shows the necessary variations in the rudder control surface deflection angle ($\delta_n$) during the test maneuver. The maximum absolute deflection angle here is 40.68°. This value is apparently acceptable in terms of geometric constraints; whereas deflections near or exceeding 60° may not be realistic, and thus suggest that the maneuver is too challenging to be accomplished with the present configuration of the airplane[373–375]. The necessary deflection angles of the elevators ($\delta_m$) and the ailerons ($\delta_l$) are much smaller than those demanded by the rudder, as shown in Fig. 15. The elevators deflection angle is positive most of the time, with negative values encountered for a brief period near the middle of the maneuver. None of the rudder deflection angle and the ailerons deflection angle are exactly symmetric about the horizontal zero line. The rudder deflection angle is positively biased, with a positive mean value of 1.294°; while the ailerons deflection angle is negatively biased, with a negative mean value of −0.624°.

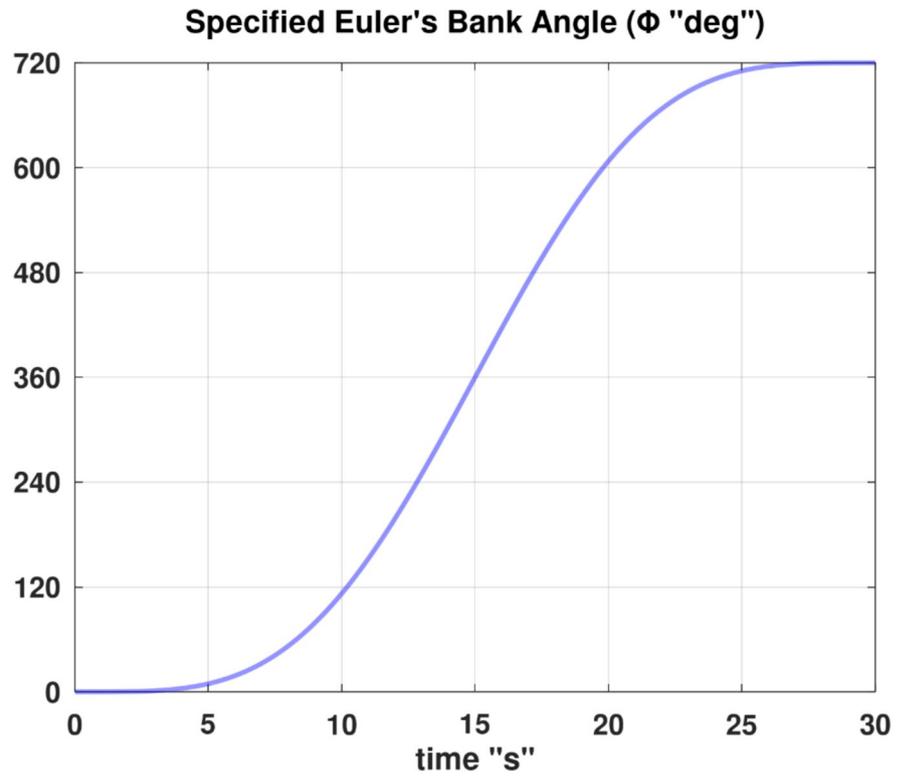

**Fig. 10**. Profile of the roll angle in the test maneuver.

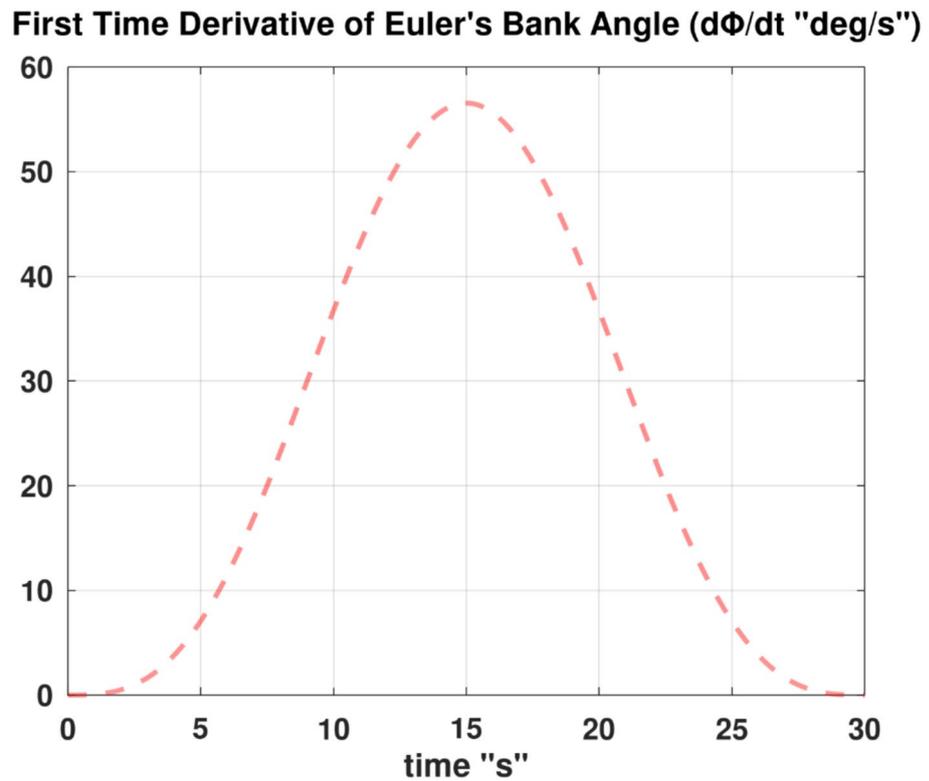

**Fig. 11**. Profile of the first time derivative of the roll angle in the test maneuver.

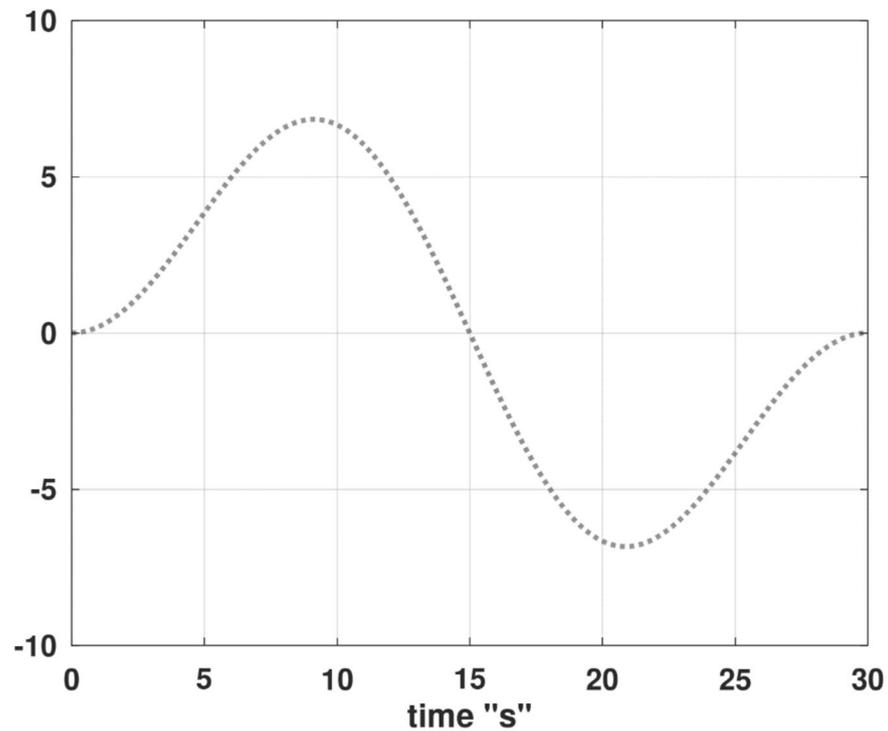

**Fig. 12.** Profile of the second time derivative of the roll angle in the test maneuver.

| Serial number | Fixed-value variable | Value |
|---|---|---|
| 1 | Flight speed ($V$) | 150 m/s (540 km/h, 291.577 knots) |
| 2 | Azimuth flight path angle ($\psi_w$) | 0° |
| 3 | Elevation flight path angle ($\theta_w$) | 0° |
| 4 | Altitude ($h$) | 5,000 m (16,404 ft; 3.1069 mi) |
| 5 | Air density ($\rho$) | 0.73587 kg/m³ |
| 6 | Dynamic pressure ($\overline{q}$) | 8,278.56 Pa (0.08170 atm) |
| 7 | Air absolute temperature ($\Theta$) | 255.65 K (−17.50 °C) |
| 8 | Air absolute pressure ($\rho R\Theta$) | 53,992 Pa (0.53286 atm) |
| 9 | Speed of sound in air ($\alpha$) | 320.50 m/s (1,153.8 km/h ; 623.00 knots) |
| 10 | Mach number ($M$) | 0.4680 |
| 11 | Equilibrium conventional angle of attack ($\widetilde{\alpha}_{equb}$) | 6.3322° |

**Table 10.** Characteristics of the proposed test maneuver of the Mirage III fighter airplane.

Figure 16 provides additional information about the inversely simulated test maneuver, through displaying the variations in the angle of attack (and its conventional counterpart) and the sideslip angle. A strong correlation can be qualitatively observed between the sideslip angle and the rudder deflection angle. This can be explained by noting that a positive sideslip angle (when the thrust vector is collinear with the longitudinal body axis) leads to a drift in the airplane's travel path toward the left (the port side) due to a continuously applied port-wise force component of the thrust, and therefore the path becomes curved. To counteract this and maintain a straight flight path, as in the target test maneuver here, a restoring moment needs to be induced through a positive rudder deflection angle (the rudder tilts toward the starboard/right side). The profile of the conventional angle of attack ($\widetilde{\alpha}$) is just a shifted version of the simulation-reported angle of attack ($\alpha$); with the difference being the frozen equilibrium conventional angle of attack ($\widetilde{\alpha}_{equb}$), which is 6.3322° (0.110518 rad) in the current test maneuver. During the maneuver, the conventional angle of attack ($\widetilde{\alpha}$) is bound between –6.1129° and 6.3322°; which is a narrow range around the zero value, thus validating the assumption of non-stall.

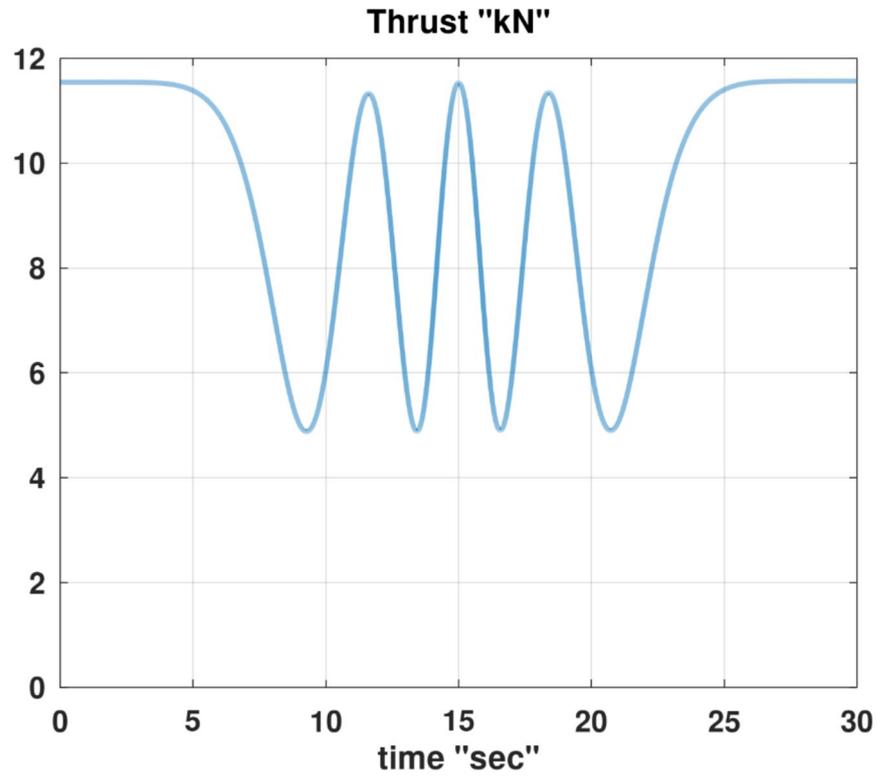

**Fig. 13.** Computed temporal profile of the thrust flight control during the test maneuver.

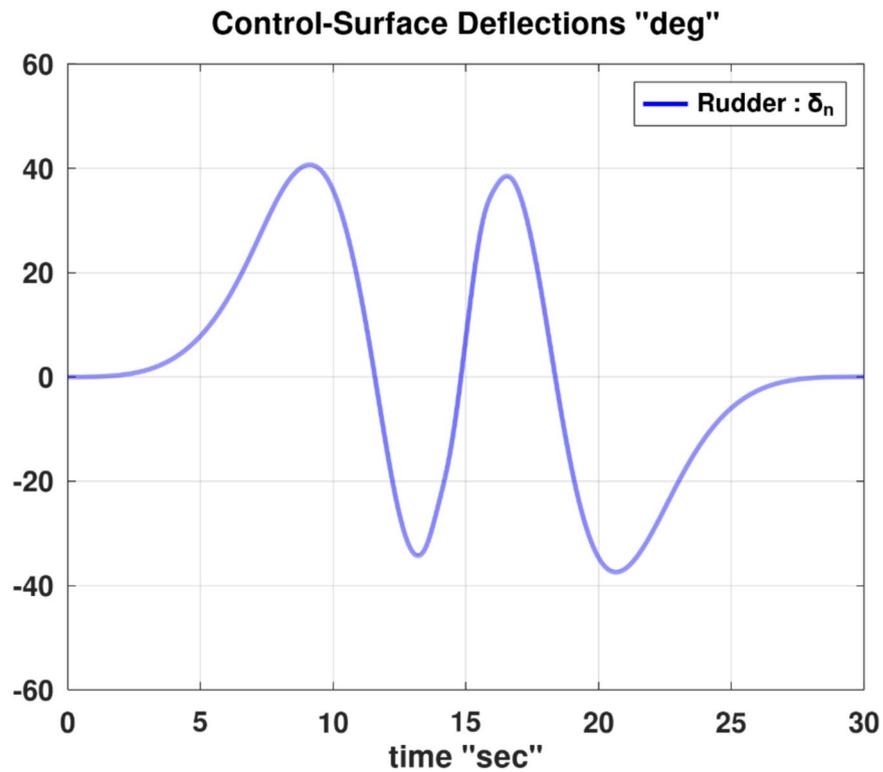

**Fig. 14.** Computed temporal profile of the rudder flight control during the test maneuver.

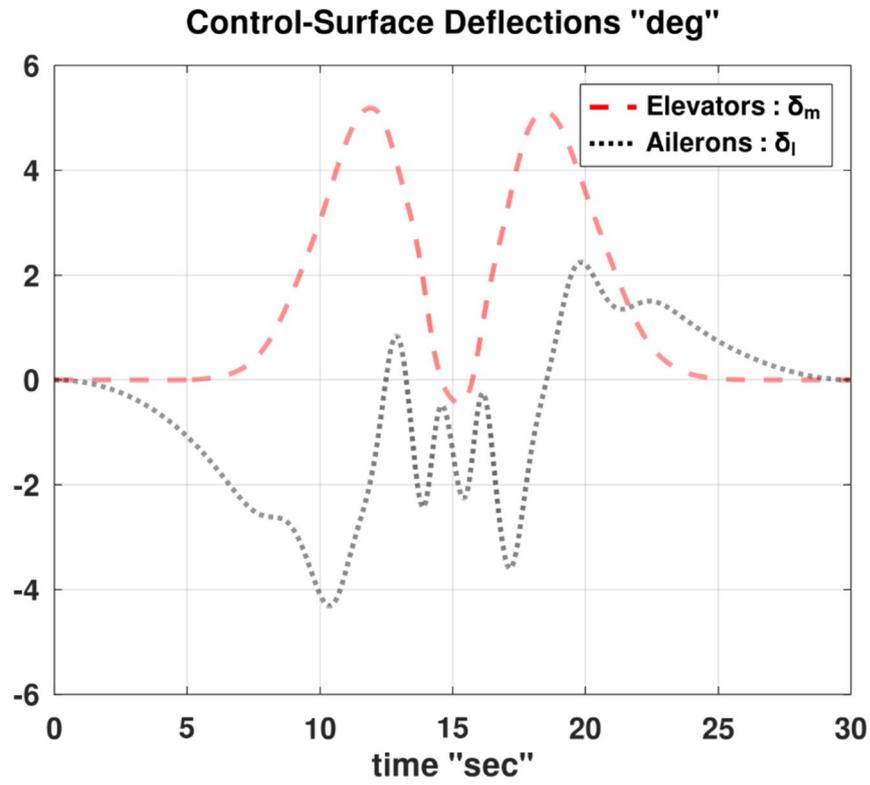

**Fig. 15**. Computed temporal profiles of the elevators and ailerons flight controls during the test maneuver.

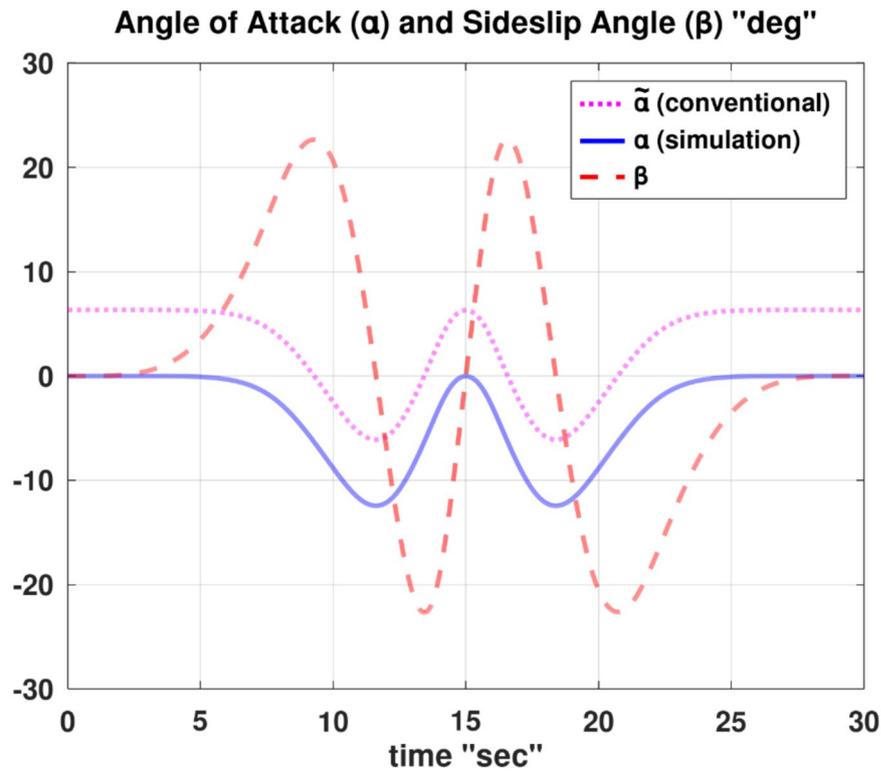

**Fig. 16**. Computed temporal profiles of the angle of attack and sideslip angle during the test maneuver.

Figure 17 provides the profiles of the two Euler angles that were not specified as input constraints, but were computed by the InvSim algorithm, and these angles are the pitch angle and the yaw (or heading) angle. Figure 18 visualizes the variation of these two airplane attitude angles, but as an orbit plot, where the pitch angle is plotted against the yaw angle with the time being a parametric variable[376,377]. This particular figure allows for judging the convergence of the simulation as the time step is successively refined. At a satisfactorily small time step (like the one used here, 0.001 s), the orbit plot shows a smooth continuous closed double-loop resembling a cardioid. In fact, this plot shows two double-loop curves on top of each other, each one is formed due to one rolling revolution by the airplane during one-half of the total maneuver duration. At an improper coarser time step (such as 0.01 s, as shown in Fig. 19; and such as 0.02 s, as shown in Fig. 20), these two curves become detached and can be visually differentiated. This means that the profiles of these two Euler angles during the first rolling revolution deviate from their profiles during the second rolling revolution, due to pronounced numerical errors.

## Discussion

In this discussion, we would like to make four supplementary comments regarding the presented study.

### Contributions

First, the contributions of the study include:

1. The detailed mathematical formulation of the three-dimensional airplane flight dynamics (the version we presented in our earlier work, which simultaneously and efficiently utilizes four different axes/coordinate systems), and its transformation from the general form to the inverse simulation (InvSim) form: This contribution included more than 80 equations, which were not simply taken from an external source. Instead, beneficial derivations and explanations were provided with the aid of visualizing sketches to properly illustrate the definition of various flight variables and the relationship among some of them. The use of manual and software-based mathematical manipulation is a valuable element of that contribution, and readers unfamiliar with such tools may find this study valuable to them in terms of either introducing them to such tools, or providing them with the final intricate expressions we obtained, making them ready to use.

2. The detailed numerical algorithm that converts the mathematical InvSim formulation into a clear procedure that can be implemented using computer programming: A large number of details, even small ones such as the quantification of errors in the gravitational acceleration at different altitudes, were presented sufficiently clear such that readers having no prior background these diverse concepts (aeronautics, atmospheric modeling, gas dynamics, thermodynamics) can deal with them without the need to go to other resources. The current study is considered self-contained to a large extent. Furthermore, many references were deliberately selected and used in this study, both as supporting points of reference and as recommended resources for interested readers to use in case they want to gain additional knowledge about a specific topic among those covered here. Even some parts of the study itself were repeated when discussed again, which effectively boosts the flow of the content.

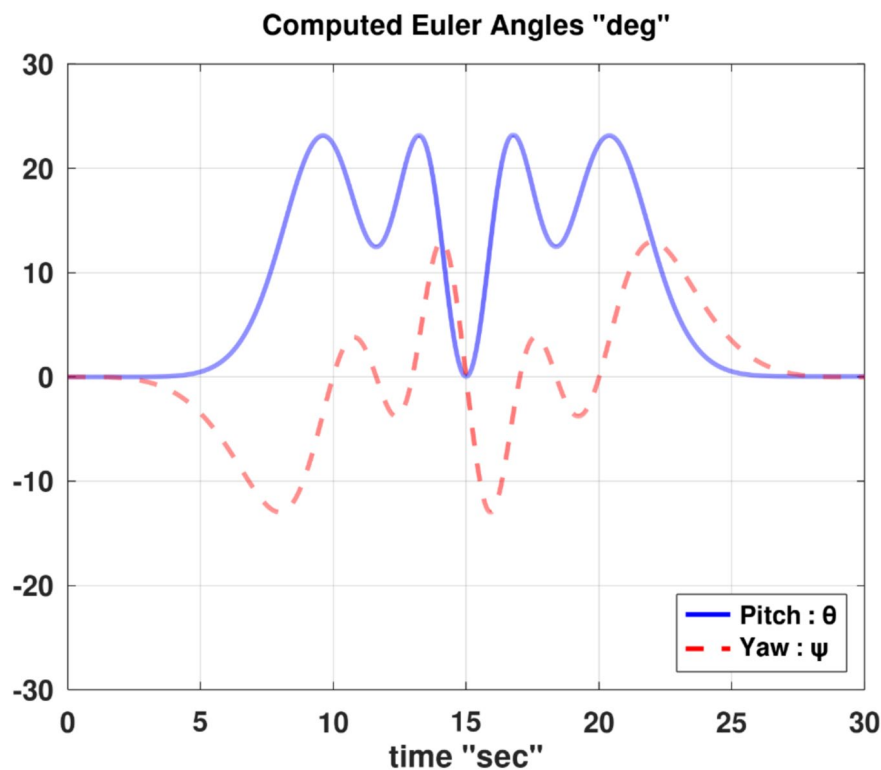

**Fig. 17.** Computed temporal profiles of the pitch and yaw angles during the test maneuver.

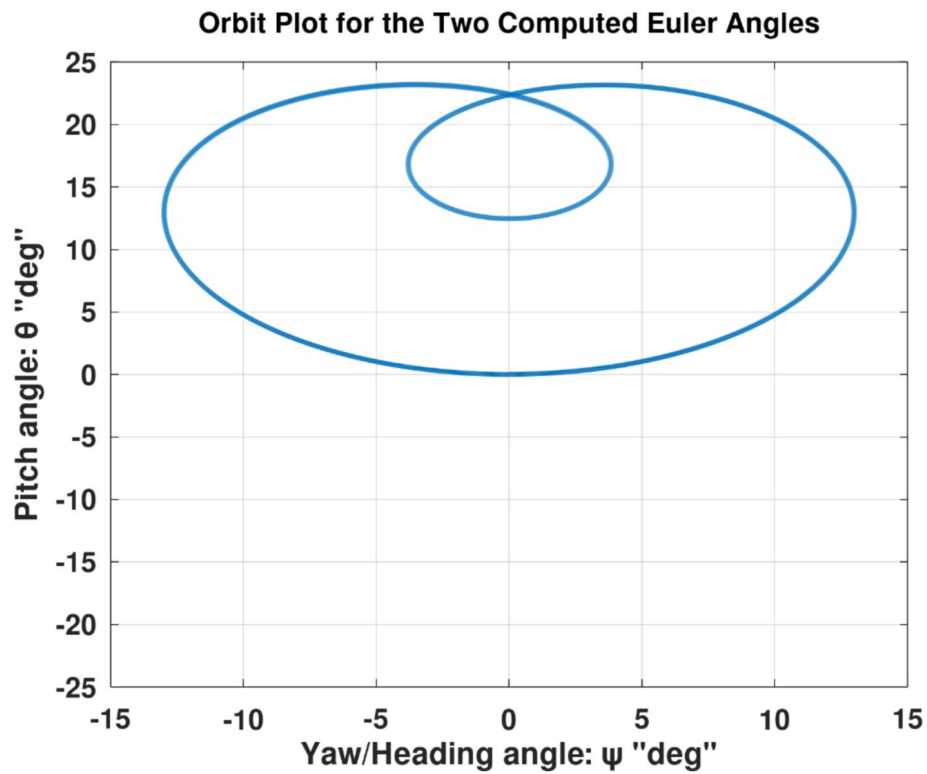

**Fig. 18**. Computed orbit plot of the pitch and yaw angles during the test maneuver at the used time step of 0.001 s.

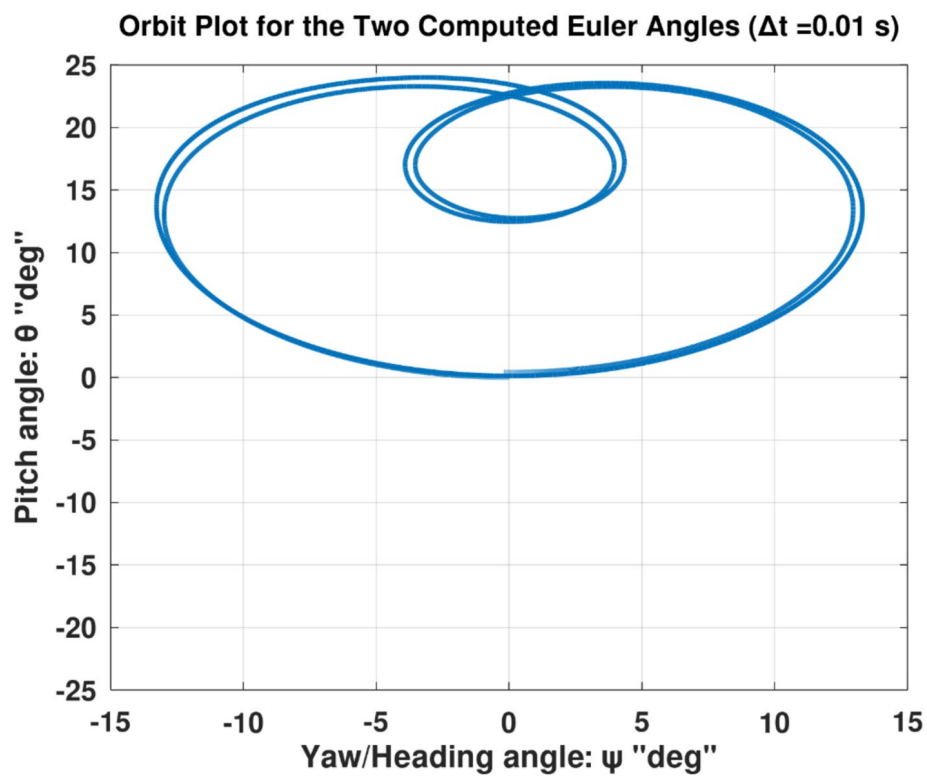

**Fig. 19**. Computed orbit plot of the pitch and yaw angles during the test maneuver at a coarse time step of 0.01 s.

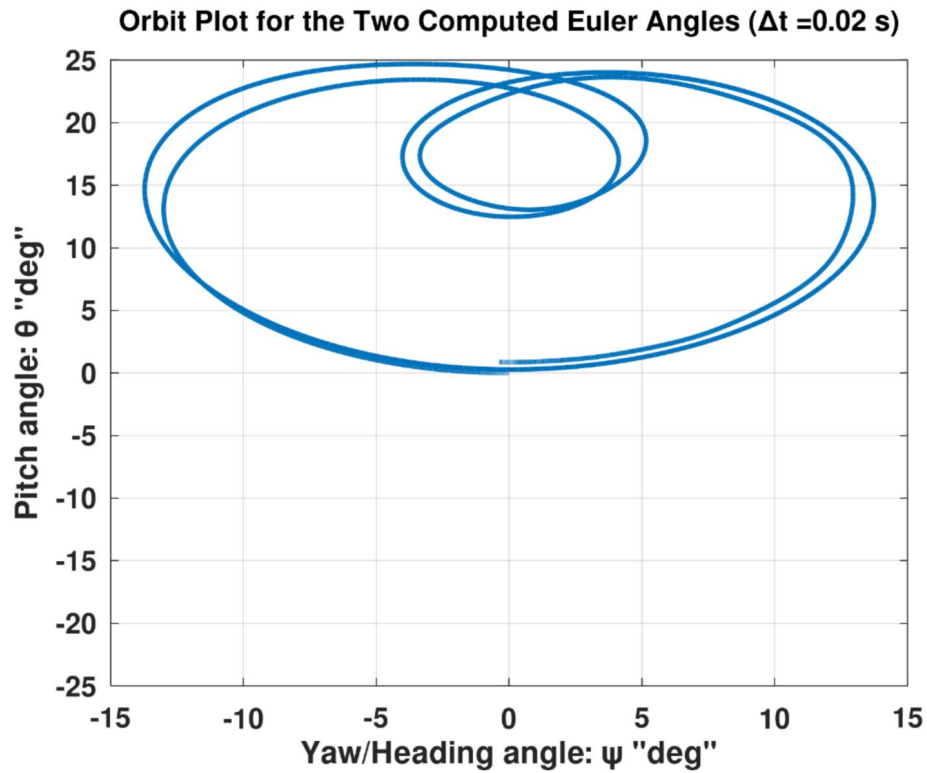

**Fig. 20.** Computed orbit plot of the pitch and yaw angles during the test maneuver at a coarser time step of 0.02 s.

3. The proposed Mirage III maneuver: This test case for flight mechanics is supplied with enough parameters that allow others to simulate the flight mission. This selected maneuver has several advantages, making it a suitable candidate for adoption as a global numerical experiment to assess other flight mechanics methods. The results of the test case presented here can also serve as a guiding benchmarking solution to compare with.

4. The detailed mathematical formulation provided here, along with closed-form expressions in multiple axes systems aided by carefully-designed illustrative sketches make this study a valuable educational asset as well for students and educators in undergraduate studies; where the study may be used partly or fully as a course content in different subjects, such as dynamics, flight mechanics, numerical methods, mathematical modeling, differential equations, robotics and artificial intelligence, and aerodynamics[378–382].

## Limitations
Second, it is admitted that the study has some limitations. We try to provide some justification for two of them below. In addition, each limitation listed below can be viewed as a potential extension in the form of another study devoted to addressing it.

1. Perhaps the main concern about this study is the lack of validation with experimental or other computational results. Although we admit this, and we feel that such validation can be a valuable addition to the study, we attempt here to mitigate this matter through the lack of such comparison data for the specific maneuver selected here (which is a novel test case). Due to the scale and nature of the test problem, being a full-scale fighter airplane, having experimental data is almost impossible. Also, the computed results provided here were supplemented with supporting elements, such as the time-sensitivity analysis and the confirmation that the solution converges successfully as the time step is reduced, the absence of numerical instability, the smoothness of the solution, and the logical pattern of the solution being a double-roll steady-level flight. In addition, the lack of validation with external sources does not affect the contributions made in this study through the mathematical and numerical analysis, the derivation of various equations, procedural steps listed toward solving the InvSim problem, and the exact list of the involved equations, variables, and constants needed. This limitation may be addressed as a separate study.

2. The airplane configuration considered here neglects some features that can be present in modern airplanes, such as the retractable landing gear. Although these elements do affect the dynamics of the airplane, the simplification made in the presented InvSim model is considered a favorable attribute, making it easier to understand and implement. The model presented can be viewed as a base model, which can be customized and expanded to handle such extra components. Again, this matter may induce additional studies.

3. The algorithm presented here adopts a north-east-down (NDE) inertial non-geocentric coordinate system for describing the trajectory of the airplane. Alternatively, a non-inertial geocentric earth-centered-earth-fixed (ECEF) coordinate system may be used, but along with another earth-centered inertial (ECI) reference frame[383–386]. Therefore, the NDE choice is more straightforward. It is admitted that the adopted NDE choice and the governing equations presented do not account for the earth's curvature and rotation. This is another limitation of the presented algorithm. However, to restrain the complexity of the computation and enable the closed-form analytical expressions, this assumption is retained. The extension to accommodate these features may be conducted in separate future studies by the interested readers. To mitigate these missing features, the algorithm should be used for modeling flight missions with a limited range (such as 1,000 km; 620 statute miles; or 870 nautical miles), in which situations the influence of earth's curvature and rotation can be reasonably dropped.

## Comparison with a state-space control model

Third, it might be useful to contrast the InvSim algorithm presented here with a control-based algorithm that follows a typical time-domain state-space representation according to the linear control theory.

For a general dynamic system (including an airplane), the governing equations may for formulated in the following standardized generic form of an ordinary differential system[387]:

$$\dot{x} = \xi(x, u, t) \tag{103}$$

where $(x)$ is a vector of state variables (states)[388] that describe the dynamics of the system and whose current values depend on their previous values, $(\dot{x})$ is the time derivative of the states, $(u)$ is a vector of inputs that is specified externally to steer and influence the system, $(t)$ is the time, and $(\xi)$ is a nonlinear function.

An output vector $(y)$ can be derived from the states and the inputs through an algebraic relation that takes the following standardized form:

$$y = \eta(x, u, t) \tag{104}$$

where $(\eta)$ is another nonlinear function.

If the two above equations can be linearized such that the original dynamic system is approximated as a linear time invariant (LTI) system[389] with no nonlinear terms and without explicit dependence on the time, then these two vector nonlinear equations can be transformed into two vector linear equations as

$$\dot{x} \cong [A]\, x + [B]u \tag{105}$$

$$y \cong [C]\, x + [D]u \tag{106}$$

where $[A]$, $[B]$, $[C]$, and $[D]$ are matrices that are obtained by a linearization technique. While the elements of these matrices are numerical values, they may either remain constant (corresponding to a common equilibrium or trim condition) or be updated over time in a discrete fashion.

In relation to our study of airplanes with six degrees of freedom, the states can be flight variables (such as the velocity components and the angular velocity components), and the input vector can be a subset of the flight variables. However, the aforementioned LTI approach to the flight mechanics problem is different from the one we presented here in several ways. The InvSim approach we presented aims to solve for the control variables, rather than having them specified as known inputs (as in the LTI approach). Also, the InvSim approach we presented does not require linearization (as in the LTI approach). Instead, it retains the full nonlinearity of the original airplane's equations of motion. Furthermore, the InvSim approach we presented does not involve any matrix operations, and does not require any matrix–vector multiplication (as in the LTI approach). Therefore, the two approaches are different, and neither one can replace the other.

## Quaternion notation versus Euler angles for attitude representation

Fourth, we elaborate on the singularity problem that occurs at $\pm 90°$ pitch angle if the orientation of the airplane (or another rigid body) is related to the inertial axes system through the sequence of three Euler angles. When the airplane nose (and its body-fixed longitudinal axis) is pointing vertically up or down, parallel to the inertial gravity axis (the pitch angle thus is $\phi = \pm 90°$), then a singularity occurs, and it is sometimes referred to as a (gimbal lock) due to the analogy with a gyroscope mounted on a mechanical gimbal system in the real world[390,391]. This singularity generally occurs when the middle rotation angle in a three-angle sequence takes a particular value that makes the rotation axes of the first and third rotations parallel. When the first and third rotational axes become aligned at this singularity condition, they coincide as a combined rotational axis, and the rotation about this common axis become effectively due to the sum of attempted rotations about the first and third axes. In our study, this means that when the pitch angle is $\pm 90°$, the roll and yaw are no longer distinguishable. An attempt to perform a rotation of 2° roll with a rotation of –3° yaw is not distinguishable from an attempt to perform a rotation of –2° roll with a rotation of 1° yaw, since the sum of both rotation angles is the same for both cases (–1°). This singularity is mathematically manifested through the appearance of infinite terms due to having $\sin(\phi)$ in the denominator; particularly $\tan(\phi)$ and $\sec(\phi)$. For example, the following transformation equations fail at this singularity condition[392]:

$$\dot{\phi} = p + q\tan{(\theta)}\sin{(\phi)} + r\tan{(\theta)}\cos{(\phi)} \tag{107}$$

$$\dot{\psi} = r\sec{(\theta)}\cos{(\phi)} + q\sec{(\theta)}\sin{(\phi)} \tag{108}$$

To cope with this singularity issue, the algorithm presented here uses intermediate coordinate systems to avoid directly relating the airplane attitude (the body axes) to the inertial axes. Instead, there is a middle moderating axis system. The presented InvSim algorithm adopts the flight-path angles $(\theta_w, \psi_w)$ and the wind axes angles $(\alpha, \beta)$; where it bypasses the singularity problem.

There is another attitude representation that can handle the aforementioned pitch singularity (or gimbal lock), which is using the quaternion notation (rotation quaternions) rather than Euler angles[393,394]. Mathematically, a quaternion can be viewed as a hyper-complex number or a four-dimensional vector, consisting of one real component and three imaginary components. Thus, a quaternion is an extended version of the classical complex number (having also one real component but only one imaginary component). A unit quaternion is a quaternion that has a norm of exactly 1. Unit quaternions can be used to represent rotations. A unit quaternion may take the following general form[395]:

$$Q = q_r + q_{im1}\widehat{i} + q_{im2}\widehat{j} + q_{im3}\widehat{k} \tag{109}$$

subject to the normalization condition

$$q_r^2 + q_{im1}^2 + q_{im2}^2 + q_{im3}^2 = 1 \tag{110}$$

Because each rotation can be represented by two quaternions ($Q$ and $-Q$); at the singularity rotation, the singularity-free quaternion (and related Euler angles solution) can be chosen, while the other is discarded.

Despite the attractive efficiency of the quaternion approach, we prefer the presented algorithm here, which has a number of merits. For example, the presented approach in this study is intuitive and relies totally on real flight quantities. It does not resort to fictitious mathematical quantities and does not require familiarity with detailed quaternion algebra.

## Conclusions

In the current study, a numerical algorithm for solving the six-degree-of-freedom inverse simulation (InvSim) flight mechanics problem for an arbitrary fixed-wing airplane was described in detail, starting from the governing equations of motion and elementary geometric relations, and including a submodel for estimating properties of atmospheric air, and steps followed in deriving the underlying system of 35 differential–algebraic equations in 39 variables, with 30 user-defined constant parameters (as well as 8 universal constants).

The algorithm utilizes the four-step Runge–Kutta method, along with second-order finite difference formulas. By specifying the aimed three rectangular coordinates of a trajectory and the aimed profile for the roll angle during a maneuver, the algorithm computes how the four flight control variables (the thrust force; and the three deflection angles of movable control surfaces – rudder, elevators, and ailerons) should change with time to achieve this target maneuver.

The algorithm was demonstrated through a complete test case of a constant-speed two-harmonic double-roll level-flight maneuver for an airplane resembling the Mirage III French fighters. The settings and results of the test problem were discussed, and the effect of the time step size was illustrated.

The work presented here may be extended to more complex operations, for example, through implementing real-time feedback control to compensate for external disturbances (such as gusts), or through accommodating additional geometric features (such as wing flaps).

## Data availability

Data generated or analyzed during this study are included in this published article. A supplementary material containing expressions for second time derivatives that were not presented in the main body of the manuscript is provided as a document titled (suppl.pdf).

## Author contributions

## Declarations

### Competing interests

### Additional information